\documentclass{article}

\usepackage{arxiv}
\usepackage[utf8]{inputenc} 
\usepackage[T1]{fontenc}    
\usepackage{hyperref}       
\usepackage{url}            
\usepackage{booktabs}       
\usepackage{amsfonts}       
\usepackage{nicefrac}       
\usepackage{microtype}      
\usepackage{lipsum}
\usepackage{graphicx}
\usepackage{amsmath,amssymb,amsthm,amsopn,array,natbib,titling}
\usepackage{authblk}
\usepackage{Sweave}
\graphicspath{{Figures/}}

\renewcommand{\today}{\begingroup
\number \day\space  \ifcase \month \or January\or February\or March\or
April\or May\or June\or July\or August\or September\or October\or
November\or December\fi
\space  \number \year \endgroup}
\newcommand{\argmin}[1]{\underset{#1}{\mathrm{argmin}} \ }
\newcommand{\argmax}[1]{\underset{#1}{\mathrm{argmax}} \ }
\def\ISE{\mathrm{ISE}}

\def\AMISE{\mathrm{AMISE}}

\def\UCV{\mathrm{UCV}}
\def\BCV{\mathrm{BCV}}
\def\CCV{\mathrm{CCV}}
\def\TCV{\mathrm{TCV}}
\def\MCV{\mathrm{MCV}}
\def\MLCV{\mathrm{MLCV}}
\def\rth{r^{th}}
\def\Intr{\int_{\boldsymbol{\mathbb{R}}}}
\def\Sum2{\sum_{i=1}^{n}\sum_{\substack{j=1 \\ j \neq i}}^{n}}
\def\M2{\mu_{2}}
\def\RK{\mathrm{R}\left(K^{(r)}\right)}
\def\RR{\mathrm{R}}
\def\Kr{K^{(r)}}
\def\ConvKr{K^{(r)} \ast K^{(r)}}
\def\Z{\left(\frac{X_{j}-X_{i}}{h}\right)}
\def\z{\left(\frac{x-X_{i}}{h}\right)}
\def\hatf{\hat{f}_{h}^{(r)}}

\let\code=\texttt

\let\pkg=\texttt
\newcommand{\CRANpkg}[1]{\href{http://CRAN.R-project.org/package=#1}{\pkg{#1}}}%
\newcommand{\email}[1]{\href{mailto:#1}{\normalfont\texttt{#1}}}

\usepackage{color}
\usepackage{fancyvrb}

\DefineVerbatimEnvironment{Highlighting}{Verbatim}{commandchars=\\\{\}}
\usepackage{framed}
\definecolor{shadecolor}{RGB}{248,248,248}

\newlength{\csllabelwidth}
\newlength{\cslhangindent}
{\setlength{\parindent}{0pt}%
	\everypar{\setlength{\hangindent}{\cslhangindent}}\ignorespaces}%
{\par}
{
	\setlength{\parindent}{0pt}
	\ifodd #1 \everypar{\setlength{\hangindent}{\cslhangindent}}\ignorespaces\fi
	\ifnum #2 > 0
	\setlength{\parskip}{#3\baselineskip}
	\fi
}%
{}
\usepackage{calc} 

\title{Kernel Estimator and Bandwidth Selection for Density and its Derivatives: The \CRANpkg{kedd} Package}
\author{by Arsalane Chouaib Guidoum\thanks{Department of Probabilities \& Statistics.\\Faculty of Mathematics. \\University of Science and Technology Houari Boumediene.\\ BP 32 El-Alia, U.S.T.H.B, Algeria.\\
		\email{acguidoum@usthb.dz}}}
\date{\today}

\begin{document}
	
\maketitle

\def\tightlist{}	

\begin{abstract}
The \pkg{kedd} package \citep{kedd} providing additional smoothing techniques to the \texttt{R} statistical system. Although various packages on the Comprehensive \texttt{R} Archive Network (CRAN) provide functions useful to nonparametric statistics, \pkg{kedd} aims to serve as a central location for more specifically of a nonparametric functions and data sets.  The current feature set of the package can be split in four main categories: compute the convolutions and derivatives of a kernel function, compute the kernel estimators for a density of probability and its derivatives, computing the bandwidth selectors with different methods, displaying the kernel estimators and selection functions of the bandwidth. Moreover, the package follows the general \texttt{R} philosophy of working with model objects. This means that instead of merely returning, say, a kernel estimator of $\rth$ derivative of a density, many functions will return an object containing, it's functions are S3 classes (\code{S3method}). The object can then be manipulated at one’s will using various extraction, summary or plotting functions. Whenever possible, we develop a graphical user interface of the various functions of a coherent whole, to facilitate the use of this package.

\end{abstract}

\keywords{
	Kernel density derivative
	\and
	Bandwidth selections
	\and
	Cross-validation
	\and
	\texttt{R} package
}

\section{Introduction}\label{Sec0}

In statistics, the univariate kernel density estimation (KDE) is a non-parametric way to estimate the probability density function $f(x)$ of a random variable $X$, is a fundamental data smoothing problem where inferences about the population are made, based on a finite data sample. This techniques are widely used in various inference procedures such as signal processing, data mining and econometrics, see e.g., \cite{Silverman,WandandJones,Jeffrey,Wolfgangetall,Alexandre}. The kernel estimator are standard in many books with applications and computer vision, see \cite{Wolfgang,Scott1992,Bowman,VenablesandRipley}, for computational complexity and with implementation in \texttt{S}, for an overview. Estimation of the density derivatives also comes up in various other applications like estimation of modes and inflexion points of densities, a good list of applications which require the estimation of density derivatives can be found in \cite{Singh1977}.

There already exist a number of packages that can perform kernel density estimation in \texttt{R} (\code{density} in \texttt{R} base); see for example \pkg{KernSmooth} \citep{KernSmooth}, \pkg{sm} \citep{smarticle}, \pkg{np} \citep{np} and \pkg{feature} \citep{feature}, they exist also of functions for kernel density derivative estimation (KDDE), e.g., \code{kdde} in \pkg{ks} package \citep{ks}. We introduce in this vignette a new \texttt{R} package \CRANpkg{kedd} \citep{kedd} for use with the statistical programming environment \cite{R}, which implements smoothing techniques and computing bandwidth selectors of the $\rth$ derivative of a probability density $f(x)$ for univariate data, using several kernels functions.

\section{Convolutions and derivatives in kernels}\label{Sec1}

In non-parametric statistics, a kernel is a weighting function used in non-parametric estimation techniques. Kernels are used in kernel density estimation to estimate random variables density functions $f(x)$, or in kernel regression to estimate the conditional expectation of a random variable, see e.g., \cite{Silverman,WandandJones}. In general any functions having the following assumptions can be used as a kernel:
\begin{itemize}
  \item[(A1)] $K(x) \geq 0$ and $\Intr K(x) dx = 1$.
  \item[(A2)] Symmetric about the origin, e.g., $\Intr x K(x) dx = 0$.
  \item[(A3)] Has finite second moment, e.g., $\M2(K) = \Intr x^{2} K(x) dx < \infty$. We denote $\RR(K) = \Intr \left(K(x)\right)^{2} dx$.
\end{itemize}
If $K(x)$ is a kernel, then so is the function $\bar{K}(x)$ defined by $\bar{K}(x)=\lambda K(\lambda x)$, where $\lambda > 0$, this can be used to select a scale that is appropriate for the data. The kernel function is very important to spreading a probability mass of $1/n$, the most widely used kernel is the Gaussian of zero mean and unit variance. Some classical of kernel function $K(x;r)$ ($r$ is the maximum derivative of kernel) in \pkg{kedd} package are the following:

\begin{table}[!ht]
\centering
\begin{tabular}{rlll}
  \toprule
  Kernel & $K(x;r)$ & $\RR(K)$ & $\M2(K)$ \\
  \midrule
  Gaussian & $K(x;\infty) =\frac{1}{\sqrt{2\pi}}\exp\left(-\frac{x^{2}}{2}\right)1_{]-\infty,+\infty[}$ & $1/\left(2\sqrt{\pi}\right)$ & 1 \\
  Epanechnikov & $K(x;2)=\frac{3}{4}\left(1-x^{2}\right)1_{(|x| \leq 1)}$ & 3/5 & 1/5 \\
  Uniform & $K(x;0)=\frac{1}{2}1_{(|x| \leq 1)}$ & 1/2 & 1/3 \\
  Triangular & $K(x;1)=(1-|x|)1_{(|x| \leq 1)}$ & 2/3 & 1/6 \\
  Triweight & $K(x;6)=\frac{35}{32}\left(1-x^{2}\right)^{3} 1_{(|x| \leq 1)}$ & 350/429 & 1/9 \\
  Tricube & $K(x;9)=\frac{70}{81}\left(1-|x|^{3}\right)^{3} 1_{(|x| \leq 1)}$ & 175/247 & 35/243 \\
  Biweight & $K(x;4)=\frac{15}{16}\left(1-x^{2}\right)^{2} 1_{(|x| \leq 1)}$ & 5/7 &  1/7\\
  Cosine & $K(x;\infty)=\frac{\pi}{4}\cos\left(\frac{\pi}{2}x\right) 1_{(|x| \leq 1)}$ & $\pi^{2}/16$  & $\left(-8+\pi^2\right)/\pi^{2}$\\
  \bottomrule
\end{tabular}
\caption{Kernel functions in \pkg{kedd} pakage.}\label{Sec1:Tab1}
\end{table}

\begin{figure}[h!]
\begin{center}
\includegraphics{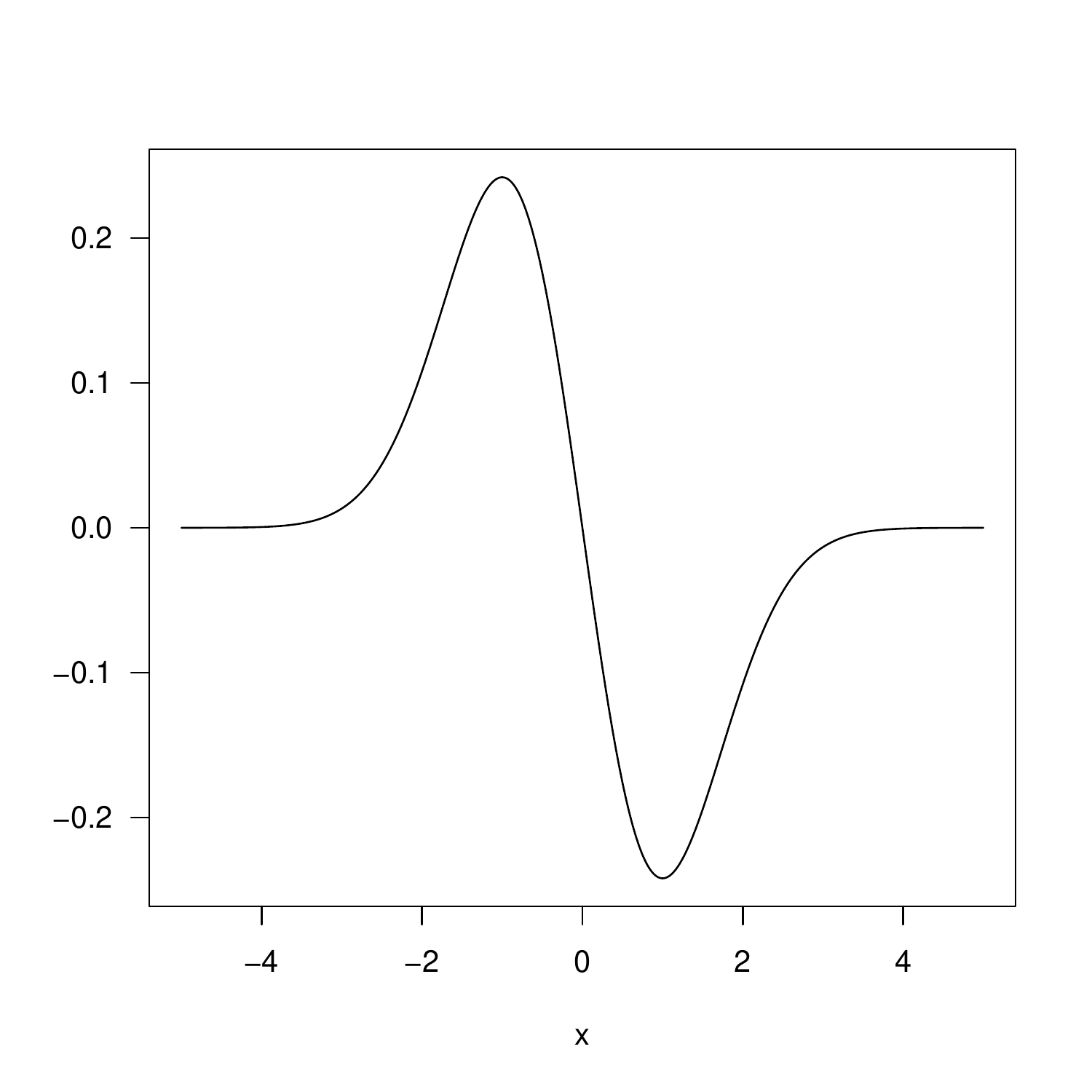}
\includegraphics{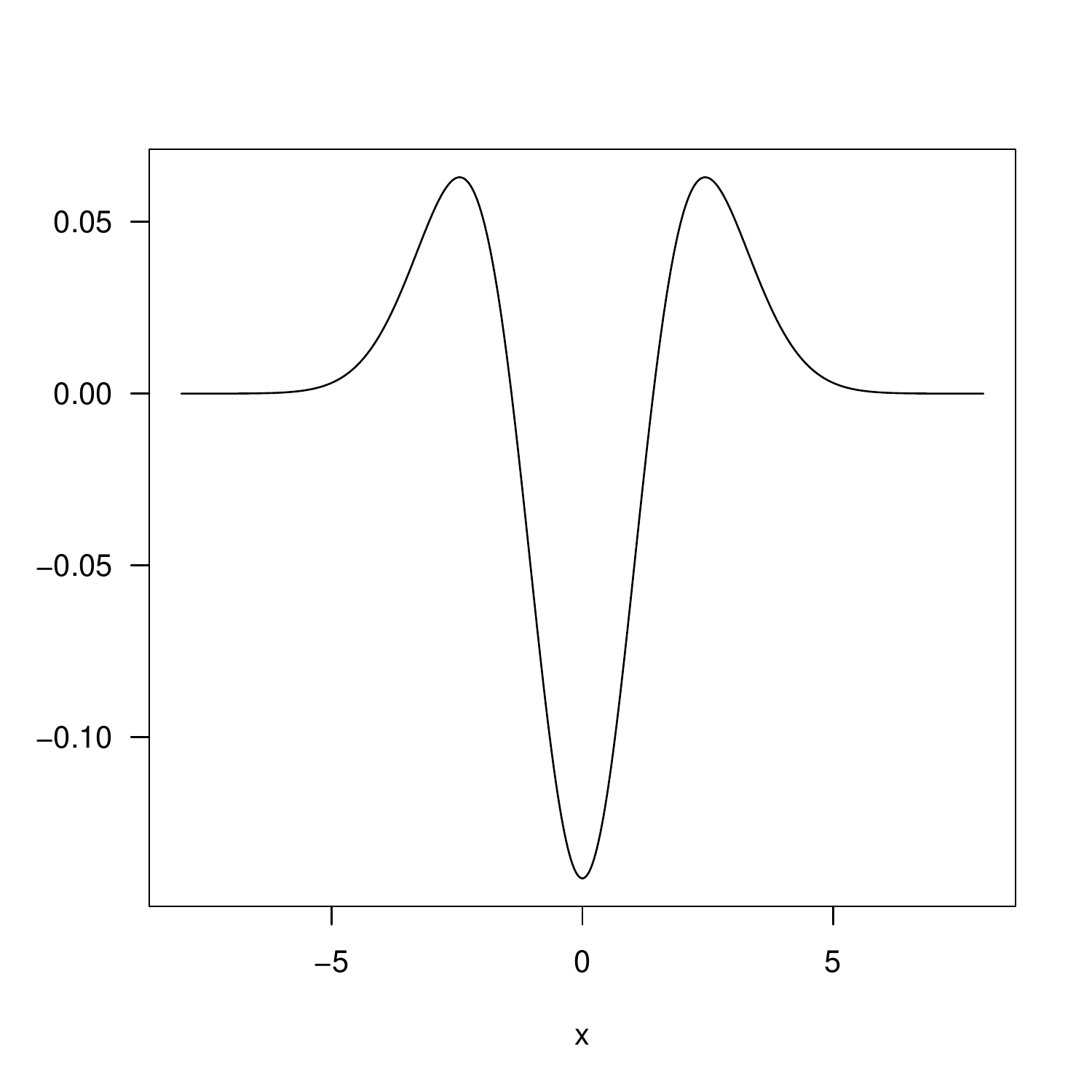}
\end{center}
\caption{(Left) First derivative of the Gaussian kernel. (Right) Convolution of the first derivative Gaussian kernel.}\label{Sec1:fig1}
\end{figure}

The $\rth$ derivative of the kernel function $K(x)$ is written as:
\begin{equation}\label{Sec1:eq1}
  \Kr(x) = \frac{d^{r}}{dx^{r}} K(x)
\end{equation}
and convolution of $\Kr(x)$ is:
\begin{equation}\label{Sec1:eq2}
  \ConvKr(x) = \Intr \Kr(x) \Kr(x-y) dy
\end{equation}
for example the $\rth$ derivative of the Gaussian kernel is given by:
$$\Kr(x) = (-1)^{r} H_{r}(x) K(x)$$
and the $\rth$ convolution can be written as:
$$\ConvKr(x) = (-1)^{2r} \Intr H_{r}(x) H_{r}(x-y)K(x)K(x-y)dy$$
where $H_{r}(x)$ is the $\rth$ Hermite polynomial, see e.g., \cite{Olver}.
We use \code{kernel.fun()} for kernel derivative defined by \eqref{Sec1:eq1}, and \code{kernel.conv()} for kernel convolution defined by \eqref{Sec1:eq2}. For example the first derivative of the Gaussian kernel displayed on the left in Figure \ref{Sec1:fig1}. On the right is the first convolution of the Gaussian kernel.
\begin{Schunk}
\begin{Sinput}
R> library(kedd)
\end{Sinput}
\end{Schunk}
\begin{Schunk}
\begin{Sinput}
R> kernel.fun(x = seq(-0.02,0.02,by=0.01), deriv.order = 1, kernel = "gaussian")$kx
\end{Sinput}
\begin{Soutput}
[1]  0.007977250  0.003989223  0.000000000 -0.003989223 -0.007977250
\end{Soutput}
\begin{Sinput}
R> kernel.conv(x = seq(-0.02,0.02,by=0.01), deriv.order = 1, kernel = "gaussian")$kx
\end{Sinput}
\begin{Soutput}
[1] -0.1410051 -0.1410368 -0.1410474 -0.1410368 -0.1410051
\end{Soutput}
\end{Schunk}
\begin{Schunk}
\begin{Sinput}
R> plot(kernel.fun(deriv.order = 1, kernel = "gaussian"))
R> plot(kernel.conv(deriv.order = 1, kernel = "gaussian"))
\end{Sinput}
\end{Schunk}

\setkeys{Gin}{width=0.45\textwidth}

\section{Kernel density derivative estimator}\label{Sec2}

Let $(X_{1},X_{2},\dots,X_{n})$ be a data sample, independent and identically distributed of a continuous random variable $X$, with density function $f(x)$. If the kernel $K$ is differentiable $r$ times then a natural estimator of the $\rth$ derivative of $f(x)$ the $\rth $ derivative of the kernel estimate \citep{Bhattacharya,Schuster,Alekseev}:
\begin{equation}\label{Sec2:eq1}
\hatf(x) = \frac{d^{r}}{dx^{r}} \frac{1}{nh} \sum_{i=1}^{n} K\z = \frac{1}{nh^{r+1}} \sum_{i=1}^{n} \Kr\z
\end{equation}
where $\Kr$ is $\rth$ derivative of the kernel function $K$, which we take to be a symmetric probability density with at least $r$ non zero derivatives when estimating $f^{(r)}(x)$, and $h$ is the bandwidth, this parameter is very important that controls the degree of smoothing applied to the data.

The following assumptions on the density $f^{(r)}(x)$, the bandwidth $h$, and the kernel $K$:
\begin{itemize}
  \item[(A4)] The $(r+2)$ derivatives $f^{(r+2)}(x)$ is continuous, square integrable and ultimately monotone.
  \item[(A5)] In the asymptotic framework, as $\lim_{n \to \infty} h_{n} = 0$ and $\lim_{n \to \infty} nh_{n}^{2r+1} = \infty$, i.e., as the number of sample $n$ is increased $h$ approaches zero at a rate slower than $1/n^{2r+1}$.
  \item[(A6)] Assumptions about $K$ are introduced in the previous section.
\end{itemize}
As seen in Equation \eqref{Sec2:eq1}, when working with a kernel estimator of the $\rth$ derivative function two choices must be made:
the kernel function $K$ and the smoothing parameter or bandwidth $h$. The choice of $K$ is a problem of less importance, because $K$ is not very sensitive to the shape of estimator, and different functions that produce good results can be used. In practice, the choice of an efficient method for the computation of $h$, for an observed data sample is a crucial problem, because of the effect of the bandwidth on the shape of the corresponding estimator. If the bandwidth is small, we will obtain an under smoothed estimator, with high variability. On the contrary, if the value of $h$ is big, the resulting estimator will be over smooth and farther from the function that we are trying to estimate.

\begin{figure}[!h]
\begin{center}
\includegraphics{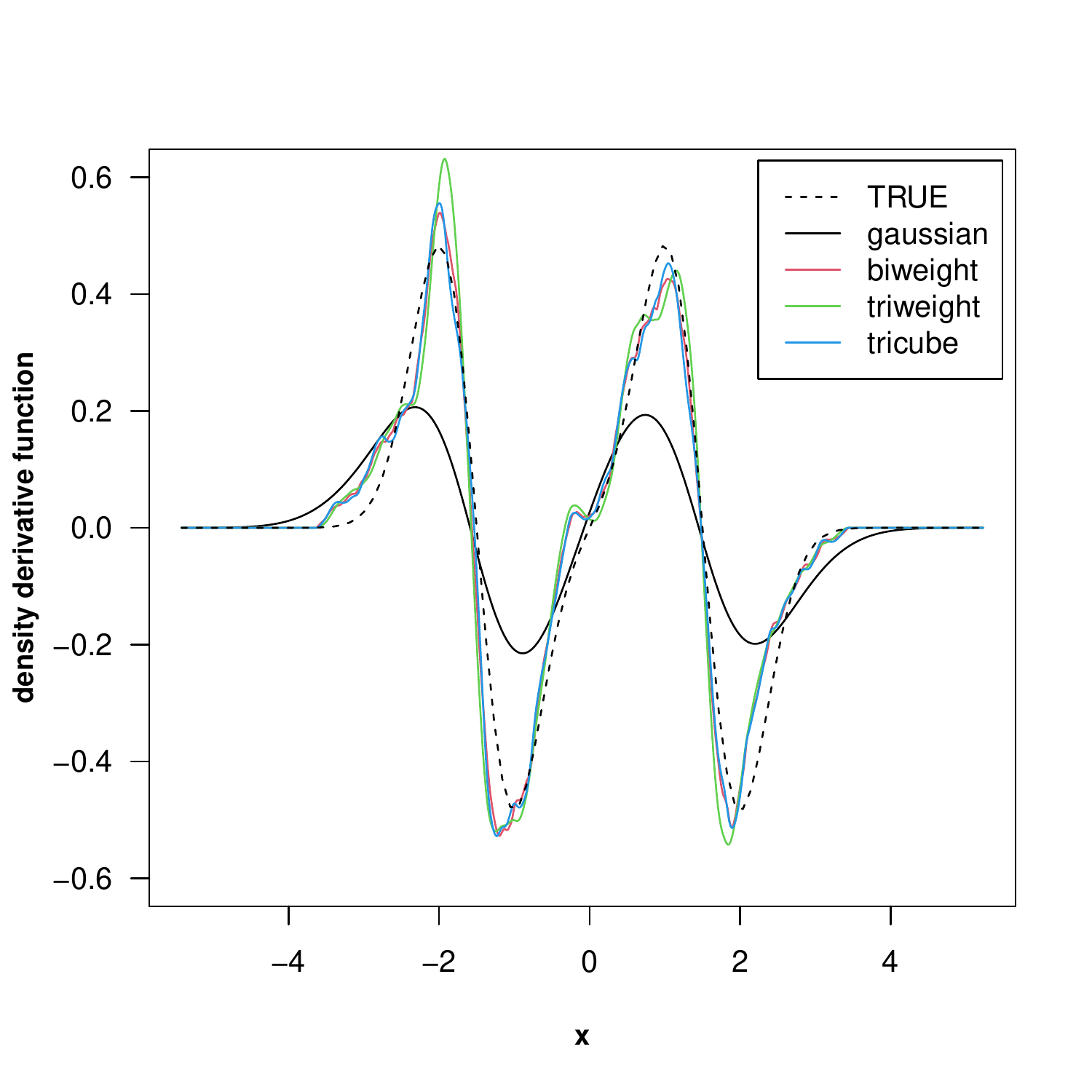}
\includegraphics{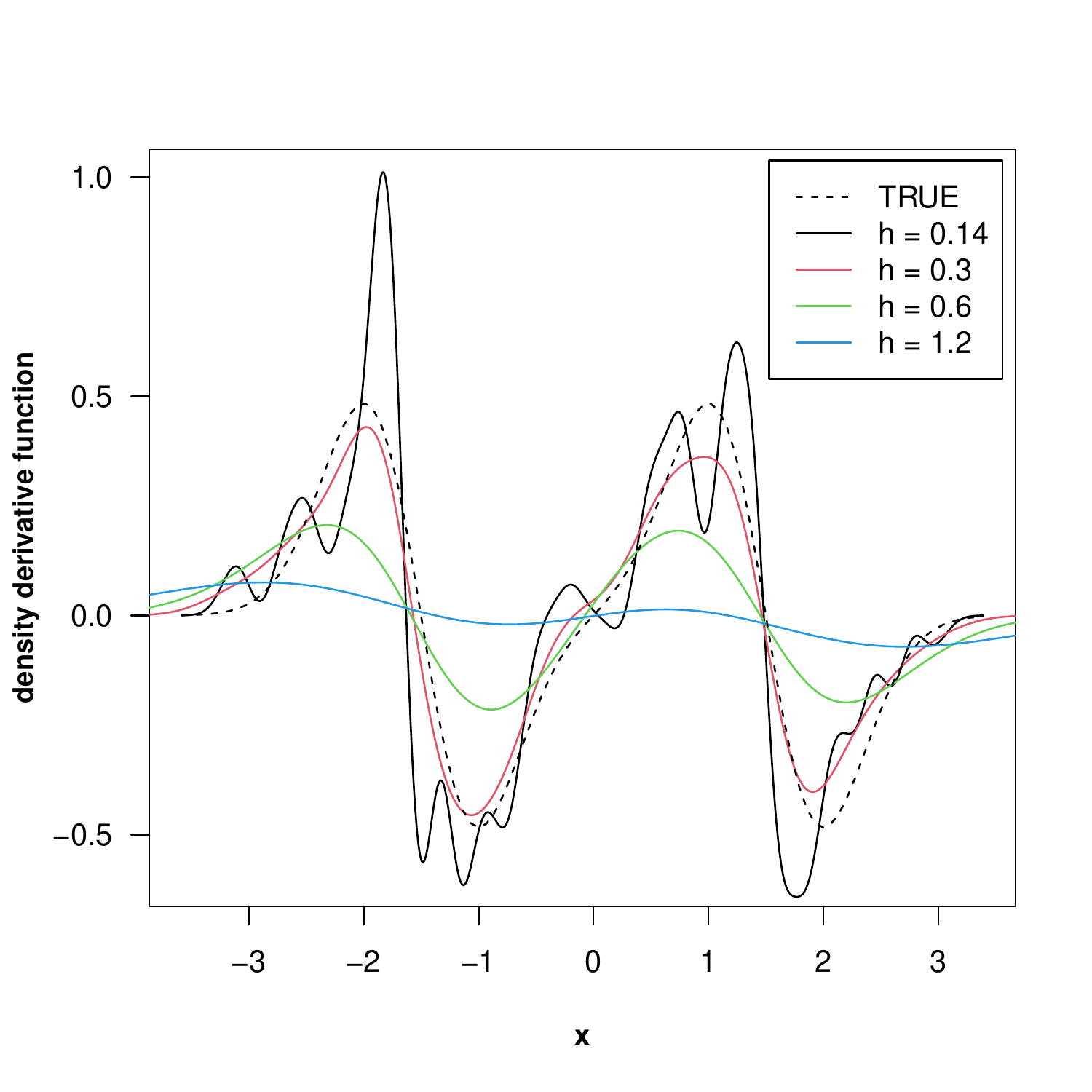}
\end{center}
\caption{(Left) Different kernels for estimation, with $h=0.6$. (Right) Effect of the bandwidth on the kernel estimator.}\label{Sec2:fig1}
\end{figure}

An example is drawn in Figure \ref{Sec2:fig1} where we show in left four different kernel (Gaussian, biweight, triweight and tricube) estimators of the first derivative of a bimodal (separated) Gaussian density (Equation \ref{Sec2:eq3}), and a given value of $h=0.6$. On the right, using the Gaussian kernel and four different values for the bandwidth.

\setkeys{Gin}{width=0.45\textwidth}

We have implemented in \texttt{R} the function \code{dkde()} corresponds to the derivative of kernel density estimator (Equation \ref{Sec2:eq1}). Eight possibilities are allowed for the kernel functions that are summarized in Table \ref{Sec1:Tab1}. We enumerate the arguments and results of this function in Table \ref{Sec2:Tab1}.

\begin{table}[!ht]
\centering	
\begin{tabular}{ll}
\toprule
  Arguments & Description \\
  \midrule
  \code{x} &  The data sample.\\
  \code{y} &  The points of the grid at which the density derivative is to be estimated.\\
           &  The default are $4h$ outside of range($x$).\\
  \code{deriv.order} & Derivative order (scalar). \\
  \code{h} &  The smoothing bandwidth to be used. The default, "ucv" unbiased cross-\\
           &  validation.\\
  \code{kernel} & The kernel function (see Table \ref{Sec1:Tab1}), by default \code{"gaussian"}. \\
   \midrule\midrule
  Results & Description \\
  \midrule\midrule
  \code{eval.points} & The coordinates of the points where the density derivative is estimated.\\
  \code{est.fx} & The estimated density derivative values (Equation \ref{Sec2:eq1}).\\
\bottomrule
\end{tabular}
\caption{Summary of arguments and results of \code{dkde()} function.}\label{Sec2:Tab1}
\end{table}

\begin{figure}[!h]
\begin{center}
\includegraphics{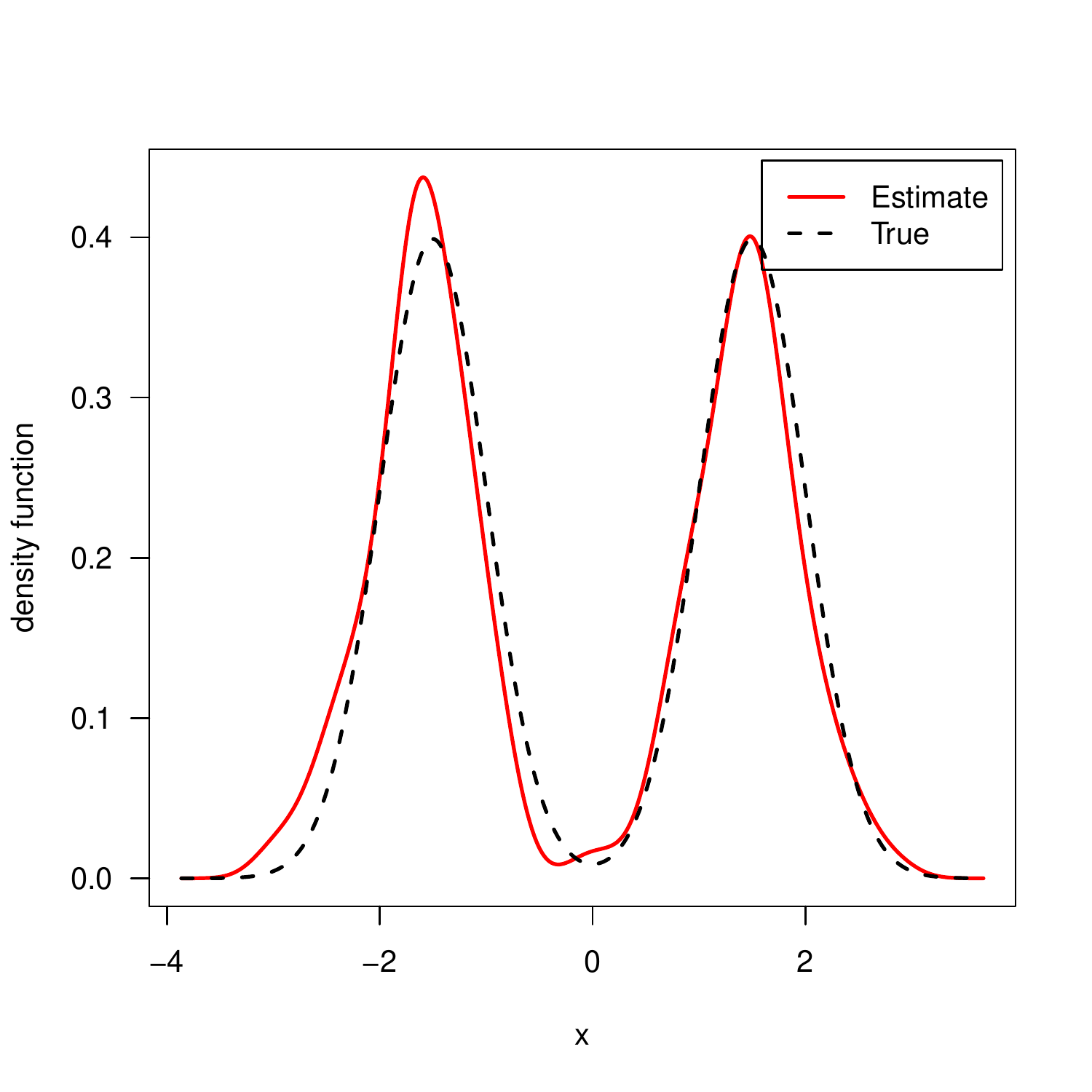}
\includegraphics{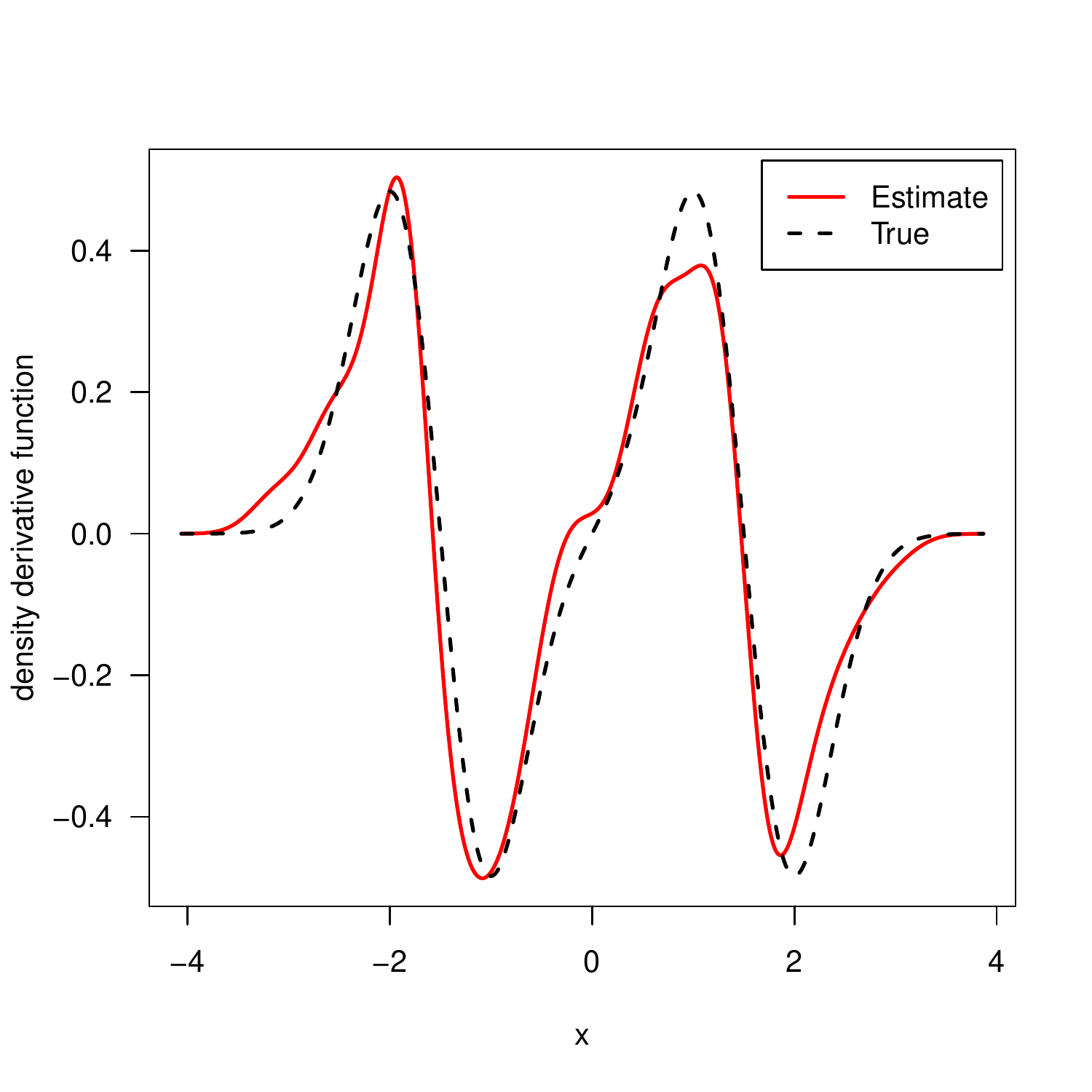}
\includegraphics{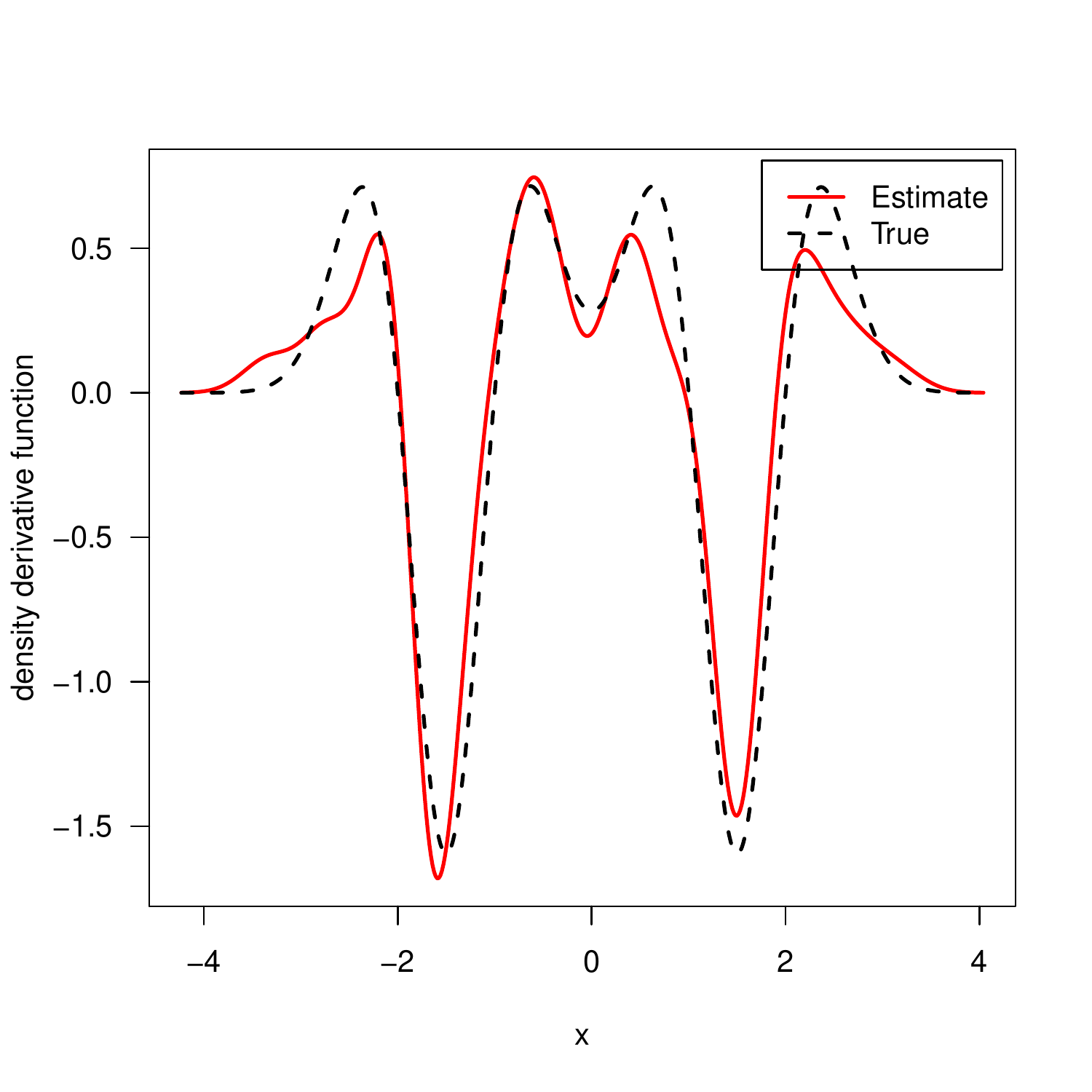}
\includegraphics{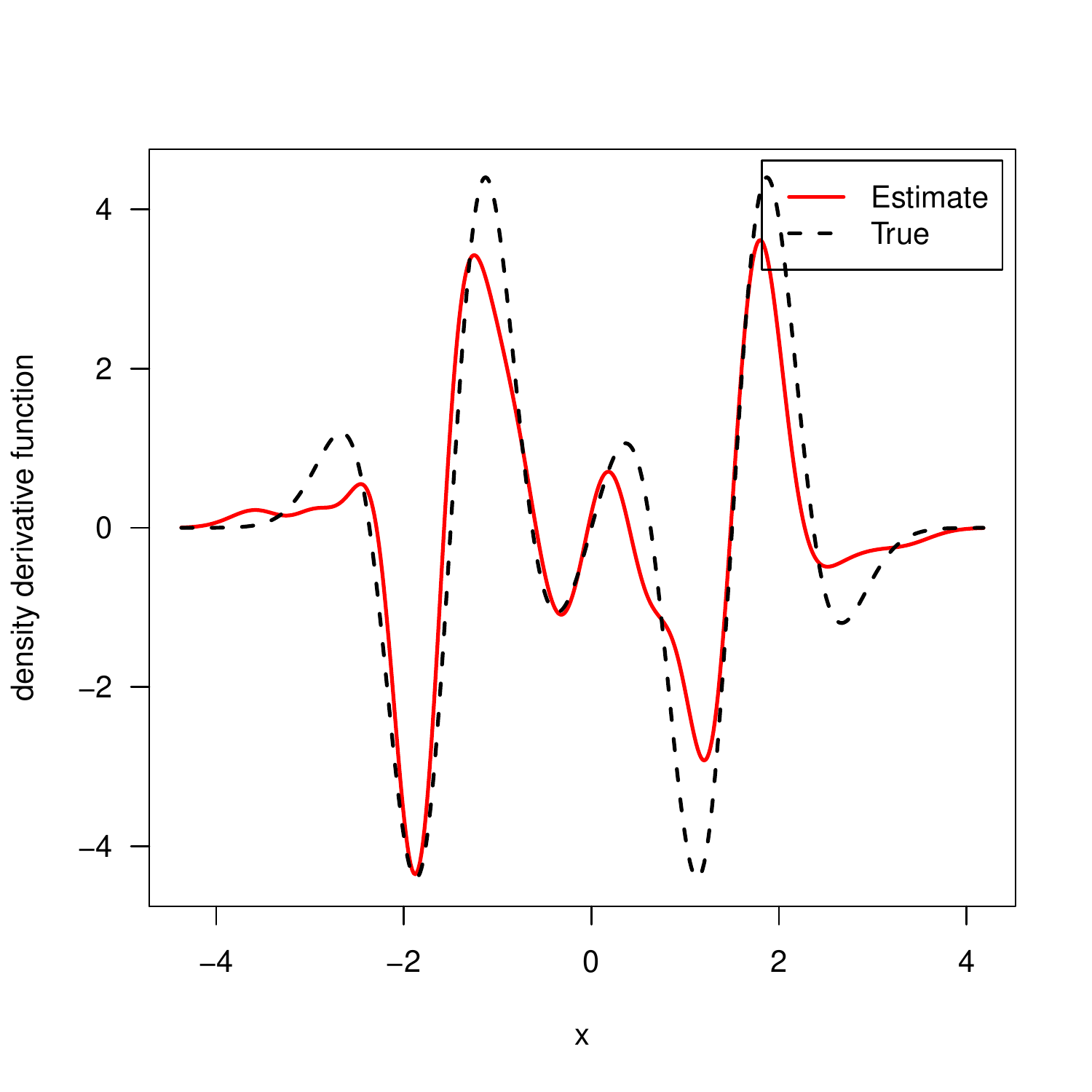}
\end{center}
\caption{Kernel density derivative estimates obtained with the function \code{dkde()}. (top left) density estimate $\hat{f}_{h}(x)$. (top right) first derivative $\hat{f}^{(1)}_{h}(x)$. (bottom left) second derivative $\hat{f}^{(2)}_{h}(x)$. (bottom right) third derivative $\hat{f}^{(3)}_{h}(x)$.}\label{Sec2:fig3}
\end{figure}

Working with the dataset \code{'bimodal'} correspond to data sample of 200 random numbers of a bi-modality (separated) of a two-component Gaussian mixture density (Equation \ref{Sec2:eq2}), with the following parameters: $-\mu_{1}=\mu_{2} = 3/2$ and $\sigma_{1}=\sigma_{2}=1/2$. The \code{dkde} function enables to compute the $\rth$ derivative of kernel density estimator over a grid of points, with a bandwidth selected by the user, but it also allows to estimate directly this parameter by the unbiased cross-validation method \code{h.ucv()} (see following Section). We have chosen this method as the automatic one because it is the fastest in computation time terms. Now we estimate the first three derivatives of $f(x)$, can be written as:
\begin{eqnarray}
  f(x) &=& 0.5\phi(\mu_{1},\sigma_{1}) + 0.5\phi(\mu_{2},\sigma_{2}) \label{Sec2:eq2}\\
  f^{(1)}(x) &=& 0.5(-4x-6) \phi(\mu_{1},\sigma_{1}) + 0.5(-4x+6)\phi(\mu_{2},\sigma_{2})\label{Sec2:eq3} \\
  f^{(2)}(x) &=& 0.5\left(\left(-4x-6\right)^{2} - 4\right)\phi(\mu_{1},\sigma_{1})+ 0.5 \left(\left(-4x+6\right)^{2} - 4\right) \phi(\mu_{2},\sigma_{2})\label{Sec2:eq4}\\
  f^{(3)}(x) &=&  0.5(-4x-6)\left(\left(-4x-6\right)^{2} - 12\right)\phi(\mu_{1},\sigma_{1})+0.5(-4x+6)  \left(\left(-4x+6\right)^{2} - 12\right)\phi(\mu_{2},\sigma_{2})\label{Sec2:eq5}
\end{eqnarray}
where $\phi$ is a standard normal density.
\begin{Schunk}
\begin{Sinput}
R> hatf  <- dkde(bimodal, deriv.order = 0)
\end{Sinput}
\begin{Soutput}
Data: bimodal (200 obs.);	Kernel: gaussian

Derivative order: 0;	Bandwidth 'h' = 0.2098

  eval.points           est.fx         
 Min.   :-3.86436   Min.   :0.0000032  
 1st Qu.:-1.98016   1st Qu.:0.0147846  
 Median :-0.09595   Median :0.0737948  
 Mean   :-0.09595   Mean   :0.1324227  
 3rd Qu.: 1.78826   3rd Qu.:0.2326044  
 Max.   : 3.67246   Max.   :0.4374314  
\end{Soutput}
\begin{Sinput}
R> hatf1 <- dkde(bimodal, deriv.order = 1)
\end{Sinput}
\begin{Soutput}
Data: bimodal (200 obs.);	Kernel: gaussian

Derivative order: 1;	Bandwidth 'h' = 0.259

  eval.points           est.fx          
 Min.   :-4.06125   Min.   :-0.4870865  
 1st Qu.:-2.07860   1st Qu.:-0.1521016  
 Median :-0.09595   Median : 0.0009041  
 Mean   :-0.09595   Mean   : 0.0000000  
 3rd Qu.: 1.88670   3rd Qu.: 0.1731795  
 Max.   : 3.86935   Max.   : 0.5038096  
\end{Soutput}
\begin{Sinput}
R> hatf2 <- dkde(bimodal, deriv.order = 2)
\end{Sinput}
\begin{Soutput}
Data: bimodal (200 obs.);	Kernel: gaussian

Derivative order: 2;	Bandwidth 'h' = 0.3017

  eval.points           est.fx          
 Min.   :-4.23200   Min.   :-1.6800486  
 1st Qu.:-2.16398   1st Qu.: 0.0012798  
 Median :-0.09595   Median : 0.1421495  
 Mean   :-0.09595   Mean   :-0.0000073  
 3rd Qu.: 1.97208   3rd Qu.: 0.3389096  
 Max.   : 4.04010   Max.   : 0.7457487  
\end{Soutput}
\begin{Sinput}
R> hatf3 <- dkde(bimodal, deriv.order = 3)
\end{Sinput}
\begin{Soutput}
Data: bimodal (200 obs.);	Kernel: gaussian

Derivative order: 3;	Bandwidth 'h' = 0.3367

  eval.points           est.fx         
 Min.   :-4.37205   Min.   :-4.353602  
 1st Qu.:-2.23400   1st Qu.:-0.472761  
 Median :-0.09595   Median : 0.001312  
 Mean   :-0.09595   Mean   :-0.000008  
 3rd Qu.: 2.04210   3rd Qu.: 0.388689  
 Max.   : 4.18016   Max.   : 3.614749  
\end{Soutput}
\end{Schunk}
By default, the function \code{dkde()} selects a grid of 512 points in the data range and used the Gaussian kernel. The output is a list containing the estimated values in the points of the grid, this last sequence and the bandwidth $h$ (by default, using unbiased cross-validation method).
In Figure \ref{Sec2:fig3} we show the first three derivatives estimators of $f(x)$ obtained with the code:
\begin{Schunk}
\begin{Sinput}
R> fx  <- function(x) 0.5 * dnorm(x,-1.5,0.5) + 0.5 * dnorm(x,1.5,0.5)
R> fx1 <- function(x) 0.5 *(-4*x-6)* dnorm(x,-1.5,0.5) + 0.5 *(-4*x+6) * 
+                    dnorm(x,1.5,0.5)
R> fx2 <- function(x) 0.5 * ((-4*x-6)^2 - 4) * dnorm(x,-1.5,0.5) + 0.5 *
+                    ((-4*x+6)^2 - 4) * dnorm(x,1.5,0.5)
R> fx3 <- function(x) 0.5 * (-4*x-6) * ((-4*x-6)^2 - 12) * dnorm(x,-1.5,0.5) +
+                      0.5 * (-4*x+6) * ((-4*x+6)^2 - 12) * dnorm(x,1.5,0.5)
R> plot(hatf ,fx = fx)
R> plot(hatf1,fx = fx1)
R> plot(hatf2,fx = fx2)
R> plot(hatf3,fx = fx3)
\end{Sinput}
\end{Schunk}
\setkeys{Gin}{width=0.45\textwidth}

\section{Bandwidth selections}\label{Sec3}

Despite the great number of bandwidth selection techniques in kernel density estimator or regression estimation, as for example \cite{Rudemo1982,Bowman1984,ScottandGeorge1987,SheatherandJones1991,Chiu1991a,Chiu1991b,Chiu1992,FeluchandKoronacki1992,Stute1992,Jonesetall1996,
Sheather2004,DuongandHazelton2003,DuongandHazelton2005,Heidenreichetall2013}, to the best of our knowledge, only few paper have been studied in the context of estimating the $\rth$ derivative of a density $f(x)$, see \citet{PeterandMarron1987,Wolfgangetall1990,JonesandKappenman1991,Stoker1993}. In this section we summarize the techniques of cross-validation methods for bandwidth choice in the kernel estimation of the derivatives of a probability density. The practicality of this methods is demonstrated by an example.

\subsection{Optimal bandwidth}

We Consider the following $\AMISE$ version of the $\rth$ derivative of a probability density $f(x)$ \cite[p. 131]{Scott1992}:
\begin{equation}\label{Sec3:eq1}
  \AMISE(h,r)= \frac{\RK}{nh^{2r+1}} + \frac{1}{4} h^{4} \M2^{2}(K) \RR\left(f^{(r+2)}\right)
\end{equation}
The optimal bandwidth minimizing \eqref{Sec3:eq1} is:
\begin{equation}\label{Sec3:eq2}
h^{\ast} = \left[\frac{(2r+1)\RK}{\M2^{2}(K) \RR\left(f^{(r+2)}\right)}\right]^{1/(2r+5)} n^{-1/(2r+5)}
\end{equation}
whereof:
\begin{equation}\label{Sec3:eq3}
   \AMISE(h,r) = \frac{2r+5}{4} \RK^{\frac{4}{(2r+5)}} \left[ \frac{\M2^{2}(K) \RR\left(f^{(r+2)}\right)}{2r+1} \right]^{\frac{2r+1}{2r+5}} n^{-\frac{4}{2r+5}}
\end{equation}
which is the smallest possible $\AMISE$ for estimation of $\hat{f}^{(r)}_{h}$. The function \code{h.amise()} provides the optimal bandwidth under $\AMISE$. The same possibilities for the kernel function as in the function \code{dkde} appear here. We enumerate the arguments and results of this function in Table \ref{Sec3:Tab1}.

\begin{table}[h!]
\centering	
\begin{tabular}{ll}
  \toprule
  Arguments & Description \\
  \midrule
  \code{x} &  The data sample.\\
  \code{deriv.order} & Derivative order (scalar). \\
  \code{lower,upper} &  Range over which to minimize. The default is almost always satisfactory,\\ 
                     &  \code{hos} (Over-smoothing) is calculated internally from an kernel.\\
  \code{tol} &  The convergence tolerance for optimize.\\ 
  \code{kernel} & The kernel function (see Table \ref{Sec1:Tab1}), by default \code{"gaussian"}. \\
  \midrule\midrule
  Results & Description \\
  \midrule\midrule
  \code{h} & Value of bandwidth (Equation \ref{Sec3:eq2}).\\
  \code{amise} & The $\AMISE$ value (Equation \ref{Sec3:eq3}).\\
  \bottomrule
\end{tabular}
\caption{Summary of arguments and results of \code{h.amise()}.}\label{Sec3:Tab1}
\end{table}

The following example computes this bandwidth for a first three derivatives estimators of \eqref{Sec2:eq2}.
\begin{Schunk}
\begin{Sinput}
R> h.amise(bimodal, deriv.order = 0)
\end{Sinput}
\begin{Soutput}
Call:		Aymptotic Mean Integrated Squared Error

Derivative order = 0
Data: bimodal (200 obs.);	Kernel: gaussian
AMISE = 0.002602521;	Bandwidth 'h' = 1.284843
\end{Soutput}
\begin{Sinput}
R> h.amise(bimodal, deriv.order = 1)
\end{Sinput}
\begin{Soutput}
Call:		Aymptotic Mean Integrated Squared Error

Derivative order = 1
Data: bimodal (200 obs.);	Kernel: gaussian
AMISE = 0.0009282042;	Bandwidth 'h' = 1.774593
\end{Soutput}
\begin{Sinput}
R> h.amise(bimodal, deriv.order = 2)
\end{Sinput}
\begin{Soutput}
Call:		Aymptotic Mean Integrated Squared Error

Derivative order = 2
Data: bimodal (200 obs.);	Kernel: gaussian
AMISE = 0.0003062873;	Bandwidth 'h' = 2.245869
\end{Soutput}
\begin{Sinput}
R> h.amise(bimodal, deriv.order = 3)
\end{Sinput}
\begin{Soutput}
Call:		Aymptotic Mean Integrated Squared Error

Derivative order = 3
Data: bimodal (200 obs.);	Kernel: gaussian
AMISE = 8.793292e-05;	Bandwidth 'h' = 2.690288
\end{Soutput}
\end{Schunk}

\subsection{Maximum likelihood cross-validation}

This method was proposed by \cite{Habbema1974} and \cite{Duin1976}. They proposed to choose $h$ so that the pseudo-likelihood $\prod_{i=1}^{n} \hat{f}_{h}(X_{i})$ is maximized. However this has a trivial maximum at $h = 0$, so the cross-validation principle is invoked by replacing $\hat{f}_{h}(x)$ by the leave-one-out $\hat{f}_{h,i}(x)$, where:
$$\hat{f}_{h,i}(X_{i}) = \frac{1}{(n-1) h} \sum_{j \neq i} K\Z$$
Define that $h$ as good which approaches the finite maximum of
\begin{equation}\label{Sec3:eq4}
  h_{mlcv} = \argmax {h > 0} \MLCV(h) 
\end{equation}
\begin{equation}\label{Sec3:eq5}
  \MLCV(h) = \left(n^{-1} \sum_{i=1}^{n} \log\left[\sum_{j \neq i} K\Z\right]-\log[(n-1)h]\right)
\end{equation}
The function \code{h.mlcv()} computed the maximum likelihood cross-validation for bandwidth selection. We enumerate the arguments and results of this function in Table \ref{Sec3:Tab2}.

\begin{table}[h!]
	\centering
\begin{tabular}{ll}
  \toprule
  Arguments & Description \\
  \midrule
  \code{x} &  The data sample.\\
  \code{lower,upper} &  Range over which to minimize. The default is almost always satisfactory.\\
  \code{tol} &  The convergence tolerance for optimize.\\
  \code{kernel} & The kernel function (see Table \ref{Sec1:Tab1}), by default \code{"gaussian"}. \\
   \midrule\midrule
  Results & Description \\
  \midrule\midrule
  \code{h} & Value of bandwidth (Equation \ref{Sec3:eq4}).\\
  \code{mlcv} & The $\MLCV$ value (Equation \ref{Sec3:eq5}).\\
  \bottomrule
\end{tabular}
\caption{Summary of arguments and results of \code{h.mlcv()}.}\label{Sec3:Tab2}
\end{table}

\begin{figure}[!ht]
\begin{center}
\includegraphics{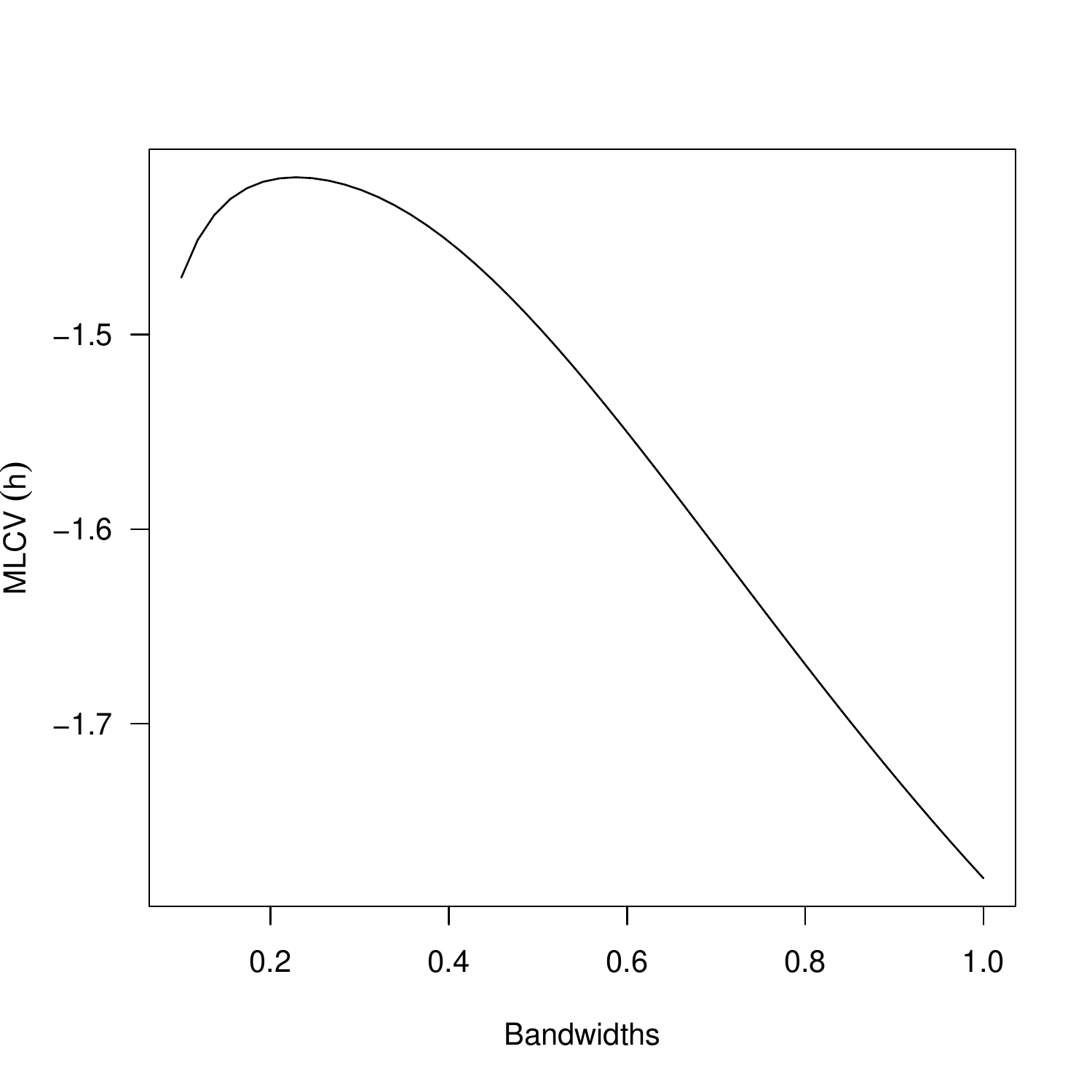}
\includegraphics{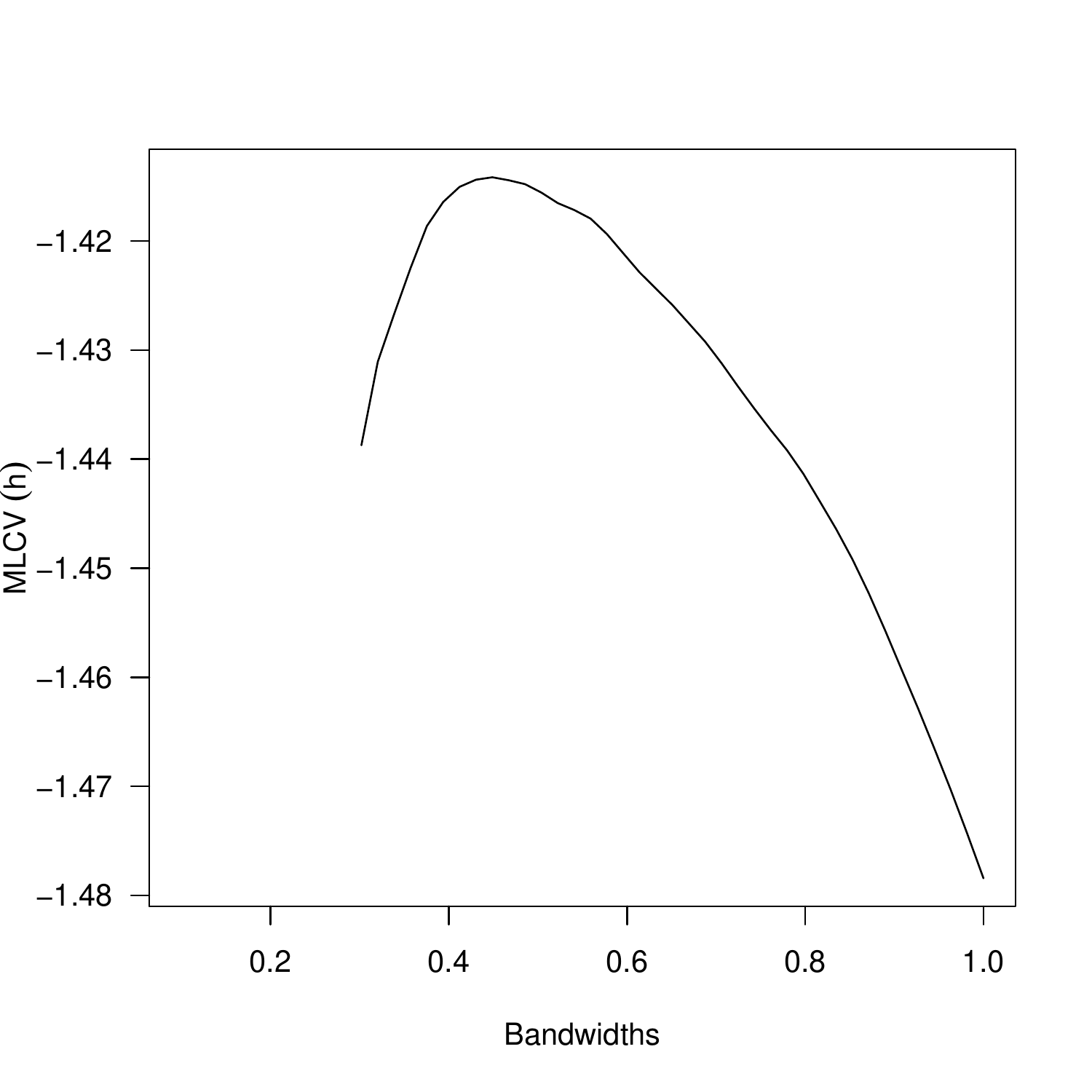}
\end{center}
\caption{$\MLCV$ function obtained by \code{h.mlcv()}, using Gaussian kernel (Left) and Epanechnikov kernel (Right).}\label{Sec3:fig1}
\end{figure}

The following example computes this bandwidth of bimodal Gaussian density (Equation \ref{Sec2:eq2}), by different kernels.
\begin{Schunk}
\begin{Sinput}
R> kernels <- eval(formals(h.mlcv.default)$kernel)
R> hmlcv <- numeric()
R> for(i in 1:length(kernels))  
+              hmlcv[i] <- h.mlcv(bimodal, kernel =  kernels[i])$h
\end{Sinput}
\end{Schunk}
\begin{Schunk}
\begin{Sinput}
R> data.frame(kernels,hmlcv)
\end{Sinput}
\begin{Soutput}
       kernels     hmlcv
1     gaussian 0.2302871
2 epanechnikov 0.4480106
3      uniform 0.3417343
4   triangular 0.4897095
5    triweight 0.6383113
6      tricube 0.5445540
7     biweight 0.5620119
8       cosine 0.4554530
\end{Soutput}
\end{Schunk}
The plot of the maximal likelihood cross validation function $\MLCV$ is shown in Figure \ref{Sec3:fig1} for Gaussian kernel in the left, and Epanechnikov kernel in the right, obtained with the code: 
\begin{Schunk}
\begin{Sinput}
R> plot(h.mlcv(bimodal, kernel =  kernels[1]), seq.bws = seq(0.1,1,length=50))
R> plot(h.mlcv(bimodal, kernel =  kernels[2]), seq.bws = seq(0.1,1,length=50))
\end{Sinput}
\end{Schunk}
\setkeys{Gin}{width=0.45\textwidth}

\subsection{Unbiased cross-validation}

\cite{Rudemo1982} and \cite{Bowman1984} proposed a so-called unbiased (least-squares) cross-validation ($\UCV$) in kernel density estimator, is probably the most popular and best studied one. An adaptation of unbiased cross-validation is proposed by \cite{Wolfgangetall1990} for bandwidth choice in the $\rth$ derivative of kernel density estimator. The essential idea of this methods, it aims to estimate $h$ the minimizer of $\ISE(h)$. The minimization criterion is defined by:
\begin{equation}\label{Sec3:eq6}
 h_{ucv} = \argmin {h > 0 }\UCV(h,r)
\end{equation}
\begin{equation}\label{Sec3:eq7}
 \UCV(h,r) = \frac{\RK}{nh^{2r+1}} + \frac{(-1)^{r}}{n (n-1) h^{2r+1}} \Sum2  \left(\ConvKr -2 K^{(2r)}\right)\Z
\end{equation}
In general, cross-validation functions in non-parametric bandwidth selection present several local minima. These minima are more likely to appear at too small values of the bandwidth \citep{PeterandMarron1991}. The function \code{h.ucv()} computes the unbiased cross-validation for bandwidth selection. We enumerate the arguments and results of this function in Table \ref{Sec3:Tab3}.

\begin{table}[!ht]
	\centering
\begin{tabular}{ll}
  \toprule
  Arguments & Description \\
  \midrule
  \code{x} &  The data sample.\\
  \code{deriv.order} & Derivative order (scalar). \\
  \code{lower,upper} &  Range over which to minimize. The default is almost always satisfactory,\\
                     &  \code{hos} (Over-smoothing) is calculated internally from an kernel.\\
  \code{tol} &  The convergence tolerance for optimize.\\
  \code{kernel} & The kernel function (see Table \ref{Sec1:Tab1}), by default \code{"gaussian"}. \\
   \midrule\midrule
  Results & Description \\
  \midrule\midrule
  \code{h} & Value of bandwidth (Equation \ref{Sec3:eq6}).\\
  \code{min.ucv} & The minimal $\UCV$ value (Equation \ref{Sec3:eq7}).\\
  \bottomrule
\end{tabular}
\caption{Summary of arguments and results of \code{h.ucv()}.}\label{Sec3:Tab3}
\end{table}

\begin{figure}[!ht]
\begin{center}
\includegraphics{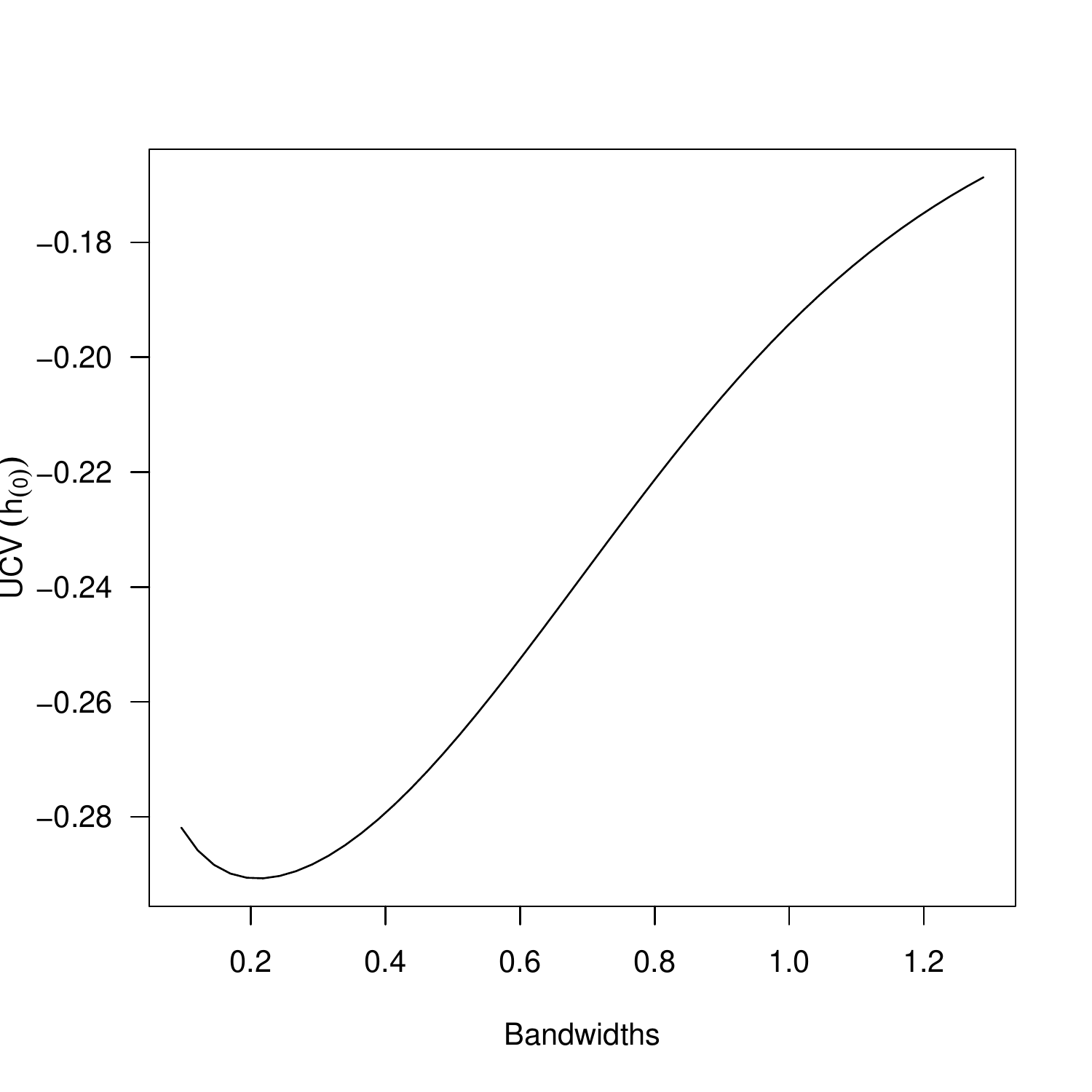}
\includegraphics{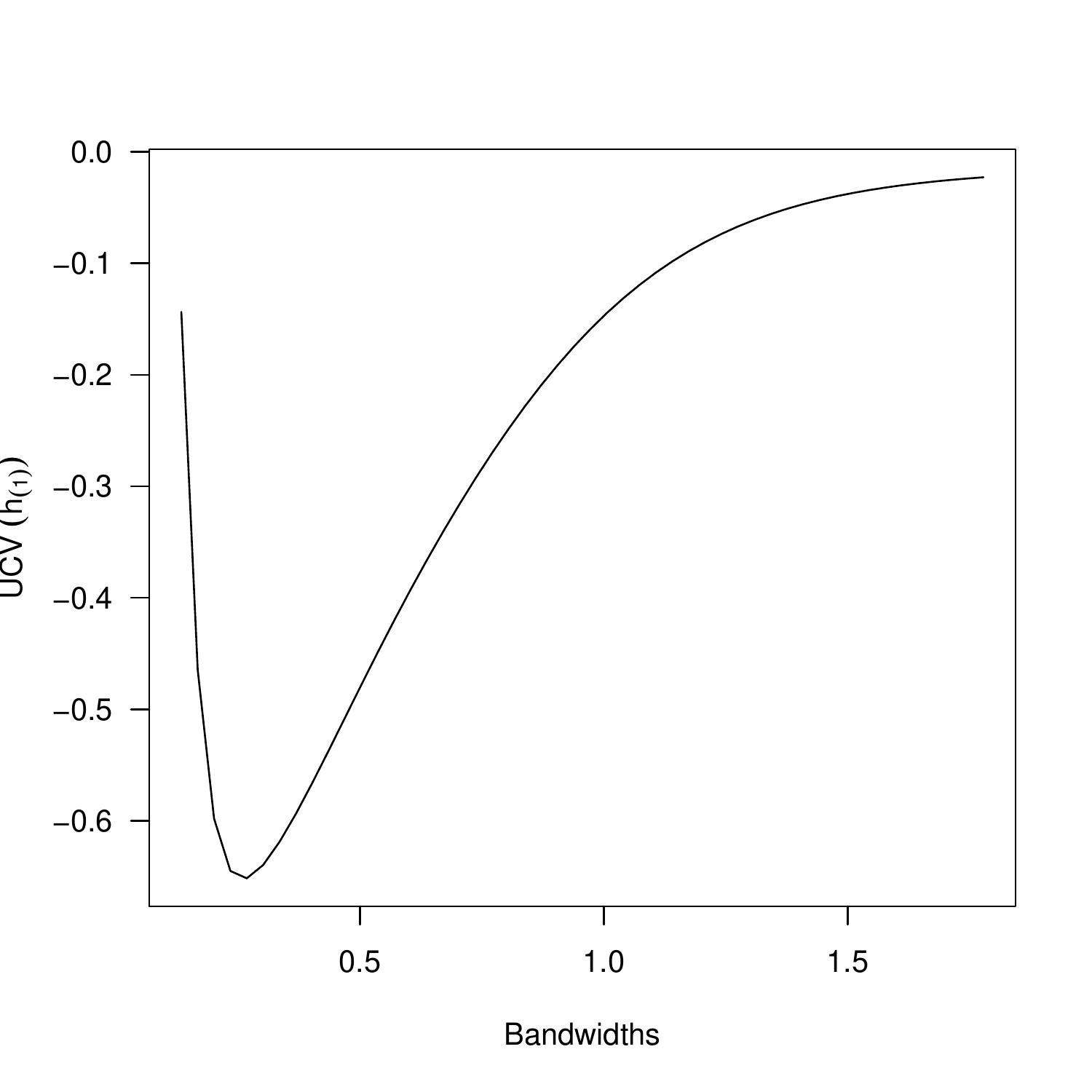}
\includegraphics{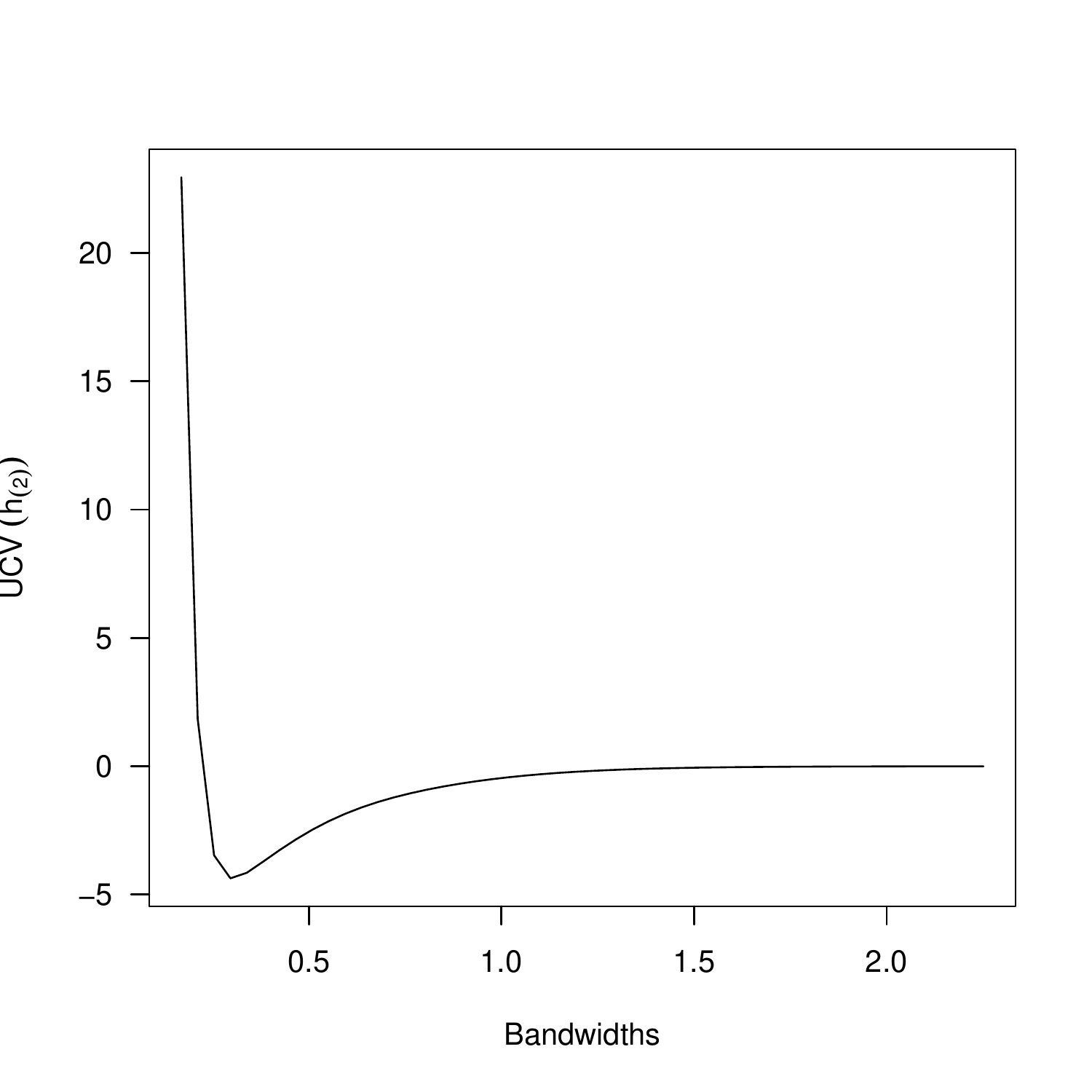}
\includegraphics{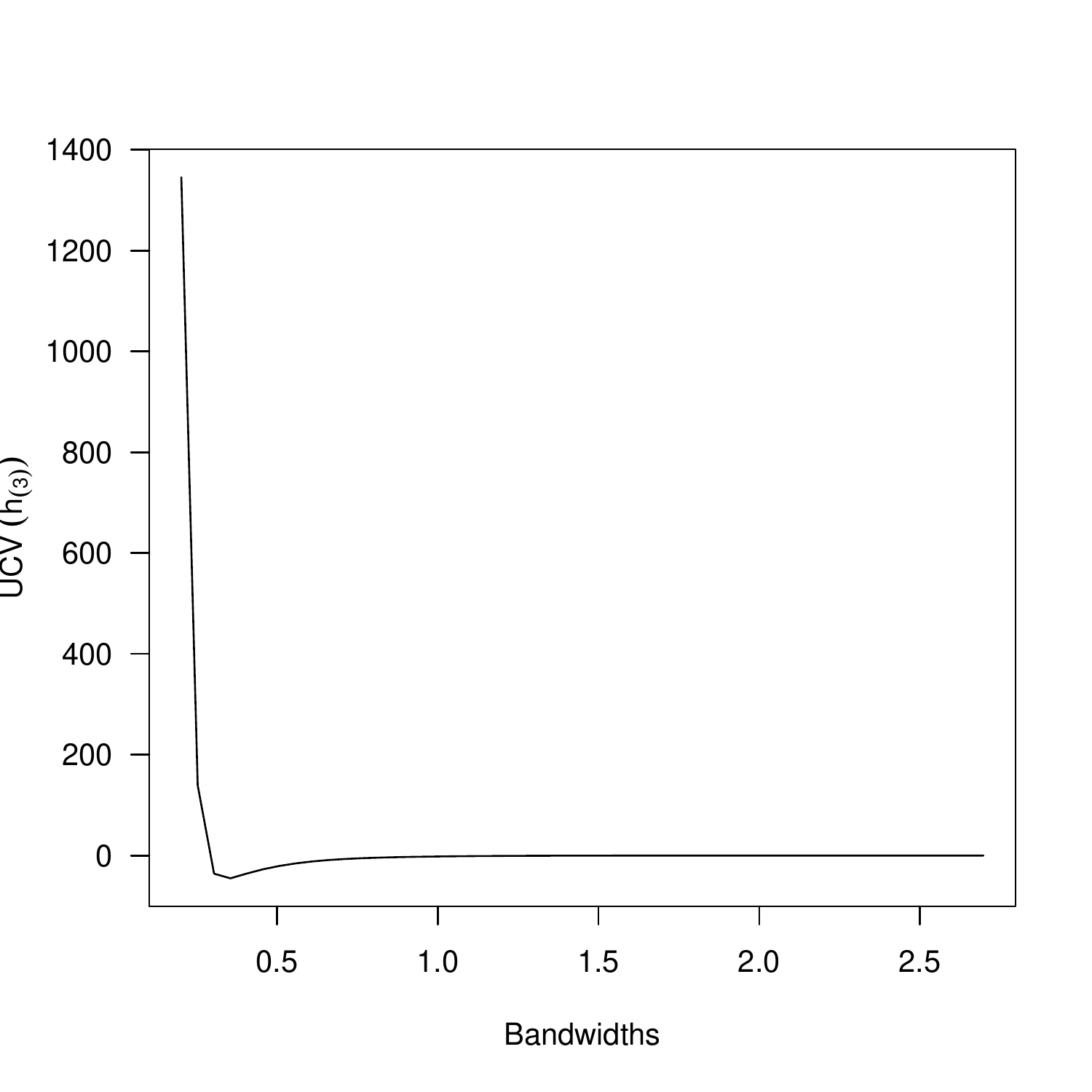}
\end{center}
\caption{$\UCV$ function obtained by \code{h.ucv()}. (top left) \code{deriv.order = 0}. (top right) \code{deriv.order = 1}. (bottom left) \code{deriv.order = 2}. (bottom right) \code{deriv.order = 3}.}\label{Sec3:fig2}
\end{figure}

The following example computes the bandwidth $h$ by this method for a first three derivatives estimators of \eqref{Sec2:eq2}.
\begin{Schunk}
\begin{Sinput}
R> h.ucv(bimodal, deriv.order = 0)
\end{Sinput}
\begin{Soutput}
Call:		Unbiased Cross-Validation

Derivative order = 0
Data: bimodal (200 obs.);	Kernel: gaussian
Min UCV = -0.290735;	Bandwidth 'h' = 0.2098
\end{Soutput}
\begin{Sinput}
R> h.ucv(bimodal, deriv.order = 1)
\end{Sinput}
\begin{Soutput}
Call:		Unbiased Cross-Validation

Derivative order = 1
Data: bimodal (200 obs.);	Kernel: gaussian
Min UCV = -0.6523005;	Bandwidth 'h' = 0.2590212
\end{Soutput}
\begin{Sinput}
R> h.ucv(bimodal, deriv.order = 2)
\end{Sinput}
\begin{Soutput}
Call:		Unbiased Cross-Validation

Derivative order = 2
Data: bimodal (200 obs.);	Kernel: gaussian
Min UCV = -4.374522;	Bandwidth 'h' = 0.3017092
\end{Soutput}
\begin{Sinput}
R> h.ucv(bimodal, deriv.order = 3)
\end{Sinput}
\begin{Soutput}
Call:		Unbiased Cross-Validation

Derivative order = 3
Data: bimodal (200 obs.);	Kernel: gaussian
Min UCV = -46.74623;	Bandwidth 'h' = 0.3367229
\end{Soutput}
\end{Schunk}
The plot of $\UCV$  function obtained with the code (Figure \ref{Sec3:fig2}):
\begin{Schunk}
\begin{Sinput}
R> for (i in 0:3) plot(h.ucv(bimodal, deriv.order = i))
\end{Sinput}
\end{Schunk}
\setkeys{Gin}{width=0.45\textwidth}

\subsection{Biased cross-validation}

Biased cross-validation was proposed by \cite{ScottandGeorge1987}, which has as its immediate target the $\AMISE$ \eqref{Sec3:eq1}. They proposed to estimate $\RR\left(f^{(r+2)}\right)$ by:
$$\hat{\RR}\left(f^{(r+2)}\right) = \RR\left(\hat{f}^{(r+2)}_{h}\right) - \frac{\RR\left(K^{(r+2)}\right)}{nh^{2r+5}}$$
There are two versions of $\BCV$, depending on the estimator of $\RR\left(f^{(r+2)}\right)$. We can use \citep{ScottandGeorge1987}
$$\hat{\RR}\left(f^{(r+2)}\right) =  \frac{(-1)^{r+2}}{n(n-1)h^{2r+5}} \Sum2 K^{(r+2)} \ast K^{(r+2)} \Z$$
or we could use \citep{JonesandKappenman1991}
$$\hat{\RR}\left(f^{(r+2)}\right) =  \frac{(-1)^{r+2}}{n(n-1) h^{2r+5}} \Sum2 K^{(2r+4)} \Z$$
From this we obtain respectively an adaptation of biased cross-validation for bandwidth choice in the $\rth$ derivative of kernel density estimator, is given by:
\begin{equation}\label{Sec3:eq8}
  \BCV_{1}(h,r) = \frac{\RK}{nh^{2r+1}} + \frac{\M2^{2}(K)}{4} \frac{(-1)^{r+2}}{n(n-1)h^{2r+1}} \Sum2 K^{(r+2)} \ast K^{(r+2)} \Z
\end{equation}
\begin{equation}\label{Sec3:eq9}
  \BCV_{2}(h,r) = \frac{\RK}{nh^{2r+1}} + \frac{\M2^{2}(K)}{4} \frac{(-1)^{r+2}}{n(n-1)h^{2r+1}} \Sum2 K^{(2r+4)} \Z
\end{equation}
The $\BCV$ selectors $h_{bcv_{1}}$ and $h_{bcv_{2}}$ are the minimisers of the appropriate $\BCV$ function. The function \code{h.bcv()} computes the biased cross-validation for bandwidth selection. We enumerate the arguments and results of this function in Table \ref{Sec3:Tab4}.

\begin{table}[!ht]
	\centering
\begin{tabular}{ll}
  \toprule
  Arguments & Description \\
  \midrule
  \code{x} &  The data sample.\\
  \code{whichbcv} & Method selected, \code{1 = BCV1} or \code{2 = BCV2}, by default \code{BCV1}. \\
  \code{deriv.order} & Derivative order (scalar). \\
  \code{lower,upper} &  Range over which to minimize. The default is almost always satisfactory,\\
                     &  \code{hos} (Over-smoothing) is calculated internally from an kernel.\\
  \code{tol} &  The convergence tolerance for optimize.\\
  \code{kernel} & The kernel function (see Table \ref{Sec1:Tab1}), by default \code{"gaussian"}. \\
  \midrule\midrule
  Results & Description \\
  \midrule\midrule
  \code{h} & Value of bandwidth.\\
  \code{min.bcv} & The minimal $\BCV$ value (Equation \ref{Sec3:eq8} or \ref{Sec3:eq9}).\\
  \bottomrule
\end{tabular}
\caption{Summary of arguments and results of \code{h.bcv()}.}\label{Sec3:Tab4}
\end{table}

\begin{figure}[!ht]
\begin{center}
\includegraphics{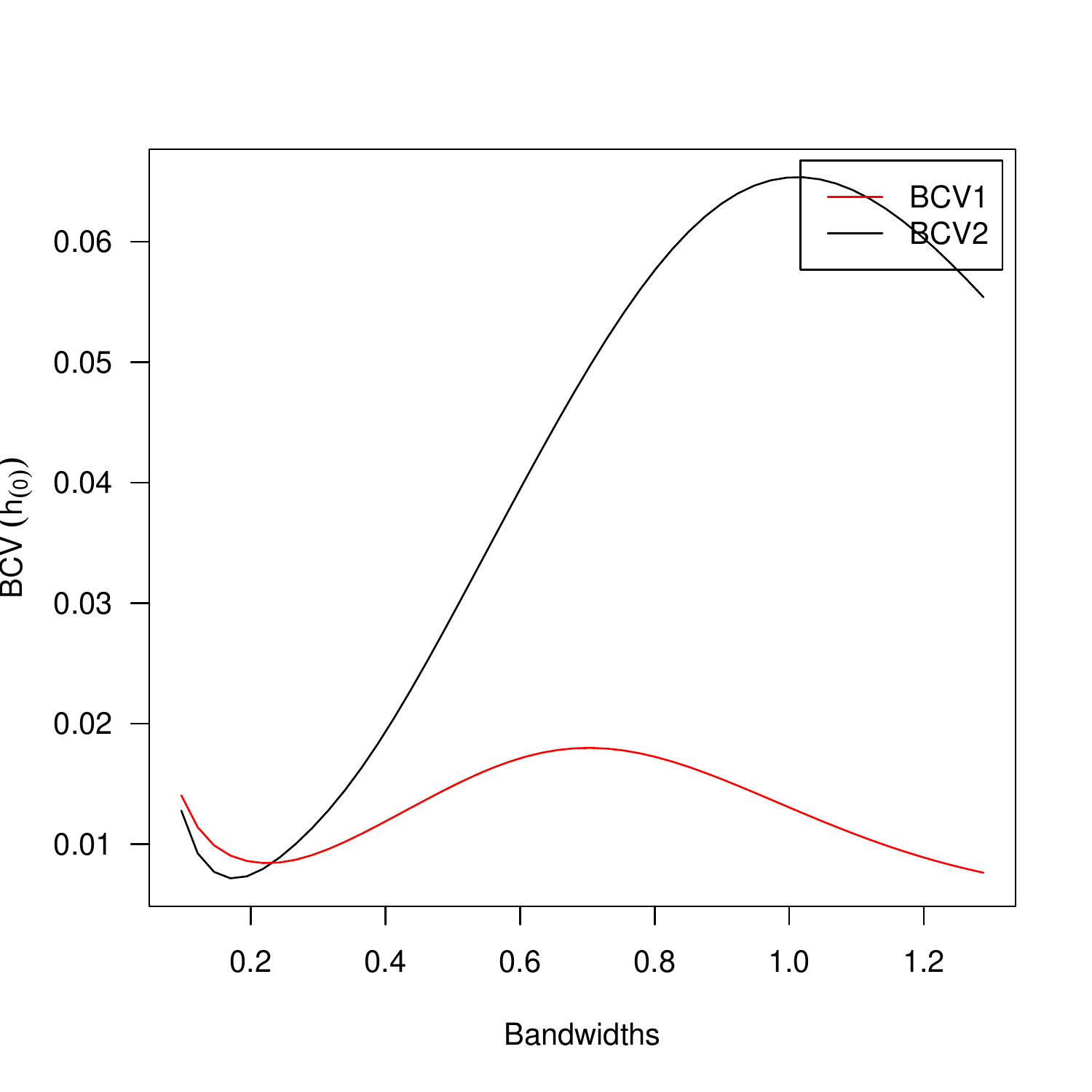}
\includegraphics{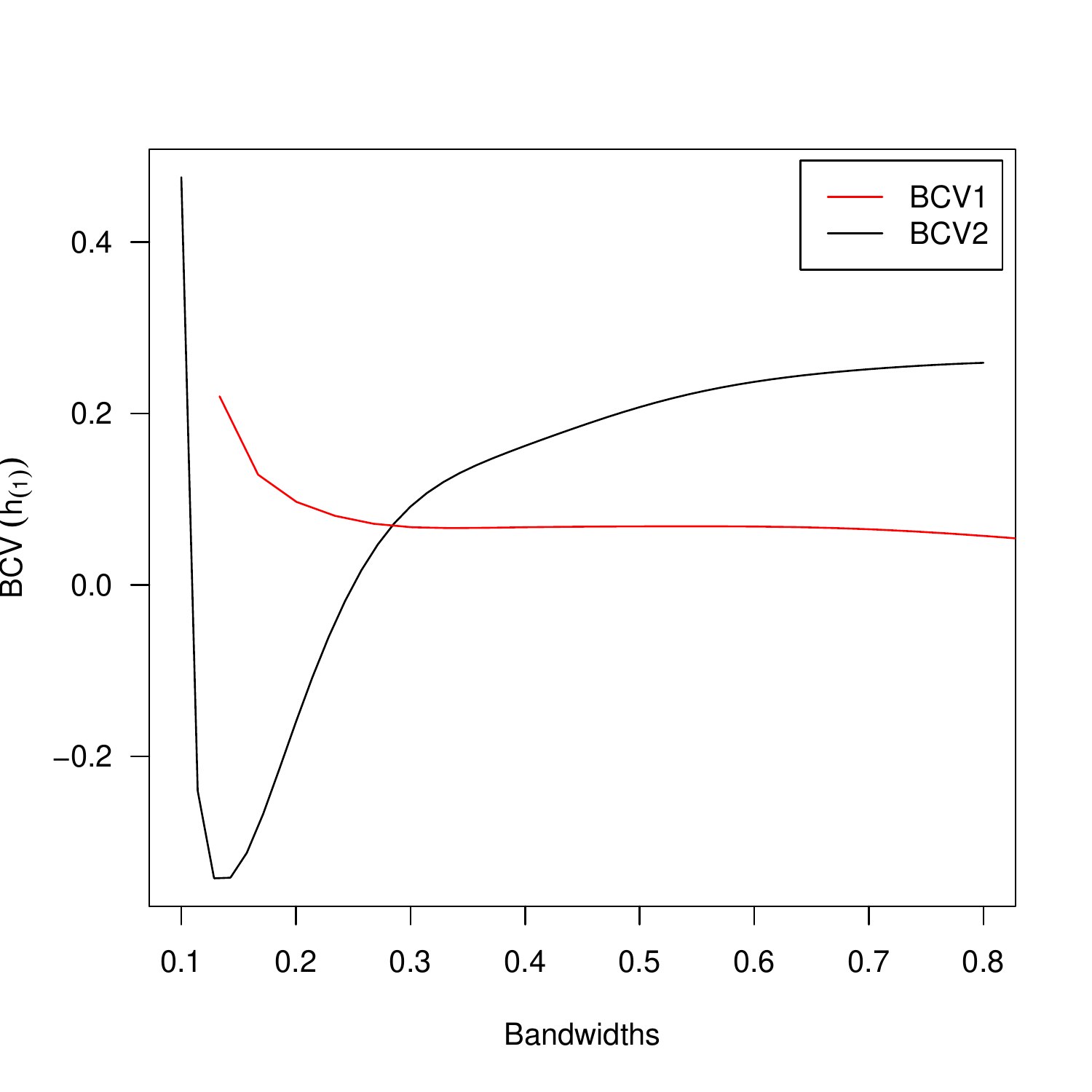}
\end{center}
\caption{$\BCV$ function obtained by \code{h.bcv()}. (Left) $\BCV_{1}$ vs $\BCV_{2}$ (\code{deriv.order = 0}). (Right) $\BCV_{1}$ vs $\BCV_{2}$ (\code{deriv.order = 1}).}\label{Sec3:fig3}
\end{figure}

The following example computes the bandwidth parameter by this method for kernel estimator of Equation \eqref{Sec2:eq2} and its first derivative estimators.
\begin{Schunk}
\begin{Sinput}
R> h.bcv(bimodal, whichbcv = 1, deriv.order = 0)
\end{Sinput}
\begin{Soutput}
Call:		Biased Cross-Validation 1

Derivative order = 0
Data: bimodal (200 obs.);	Kernel: gaussian
Min BCV = 0.008406978;	Bandwidth 'h' = 0.2239608
\end{Soutput}
\begin{Sinput}
R> h.bcv(bimodal, whichbcv = 2, deriv.order = 0)
\end{Sinput}
\begin{Soutput}
Call:		Biased Cross-Validation 2

Derivative order = 0
Data: bimodal (200 obs.);	Kernel: gaussian
Min BCV = 0.007133677;	Bandwidth 'h' = 0.1748194
\end{Soutput}
\begin{Sinput}
R> h.bcv(bimodal, whichbcv = 1, deriv.order = 1, lower=0.1, upper=0.8)
\end{Sinput}
\begin{Soutput}
Call:		Biased Cross-Validation 1

Derivative order = 1
Data: bimodal (200 obs.);	Kernel: gaussian
Min BCV = 0.06626954;	Bandwidth 'h' = 0.3366177
\end{Soutput}
\begin{Sinput}
R> h.bcv(bimodal, whichbcv = 2, deriv.order = 1, lower=0.1, upper=0.8)
\end{Sinput}
\begin{Soutput}
Call:		Biased Cross-Validation 2

Derivative order = 1
Data: bimodal (200 obs.);	Kernel: gaussian
Min BCV = -0.3470459;	Bandwidth 'h' = 0.1356763
\end{Soutput}
\end{Schunk}
The plot of $\BCV$  function obtained with the code \code{h.bcv()} (Figure \ref{Sec3:fig3}):
\begin{Schunk}
\begin{Sinput}
R> ## deriv.order = 0
R> plot(h.bcv(bimodal, whichbcv = 2, deriv.order = 0))
R> lines(h.bcv(bimodal, whichbcv = 1, deriv.order = 0),col="red")
R> legend("topright", c("BCV1","BCV2"),lty=1,col=c("red","black"),
+        inset = .015)
R> ## deriv.order = 1
R> plot(h.bcv(bimodal, whichbcv = 2, deriv.order = 1),seq.bws = 
+      seq(0.1,0.8,length=50))
R> lines(h.bcv(bimodal, whichbcv = 1, deriv.order = 1),col="red")
R> legend("topright", c("BCV1","BCV2"),lty=1,col=c("red","black"),
+       inset = .015)
\end{Sinput}
\end{Schunk}
\setkeys{Gin}{width=0.45\textwidth}

\subsection{Complete cross-validation}

\cite{JonesandKappenman1991} proposed a so-called complete cross-validation ($\CCV$) in kernel density estimator. This method can be extended to the estimation of derivative of the density, basing our estimate of integrated squared density derivative \citep{PeterandMarron1987} we get the following. Thus, $h_{ccv}$, say, is the $h$ that minimises:
\begin{equation}\label{Sec3:eq10}
    \CCV(h,r) = \RR\left(\hatf\right) -\bar{\theta}_{r}(h) + \frac{1}{2}\M2(K) h^{2} \bar{\theta}_{r+1}(h)+ \frac{1}{24}\left(6\M2^{2}(K) -\delta(K)\right) h^{4}\bar{\theta}_{r+2}(h)
\end{equation}
where,
$$\RR\left(\hatf\right) = \frac{\RK}{nh^{2r+1}} + \frac{(-1)^{r}}{n (n-1) h^{2r+1}} \Sum2 \ConvKr \Z$$
and
$$\bar{\theta}_{r}(h)= \frac{(-1)^r}{n(n-1) h^{2r+1}} \Sum2 K^{(2r)} \Z$$
with : $\delta(K) = \Intr x^{4} K(x) dx$.

The function \code{h.ccv()} computes the complete cross-validation for bandwidth selection. We enumerate the arguments and results of this function in Table \ref{Sec3:Tab5}.

\begin{table}[!ht]
	\centering
\begin{tabular}{ll}
  \toprule
  Arguments & Description \\
  \midrule
  \code{x} &  The data sample.\\
  \code{deriv.order} & Derivative order (scalar). \\
  \code{lower,upper} &  Range over which to minimize. The default is almost always satisfactory,\\
                     &  \code{hos} (Over-smoothing) is calculated internally from an kernel.\\
  \code{tol} &  The convergence tolerance for optimize.\\
  \code{kernel} & The kernel function (see Table \ref{Sec1:Tab1}), by default \code{"gaussian"}. \\
   \midrule\midrule
  Results & Description \\
  \midrule\midrule
  \code{h} & Value of bandwidth.\\
  \code{min.ccv} & The minimal $\CCV$ value (Equation \ref{Sec3:eq10}).\\
  \bottomrule
\end{tabular}
\caption{Summary of arguments and results of \code{h.ccv()}.}\label{Sec3:Tab5}
\end{table}

\begin{figure}[!ht]
\begin{center}
\includegraphics{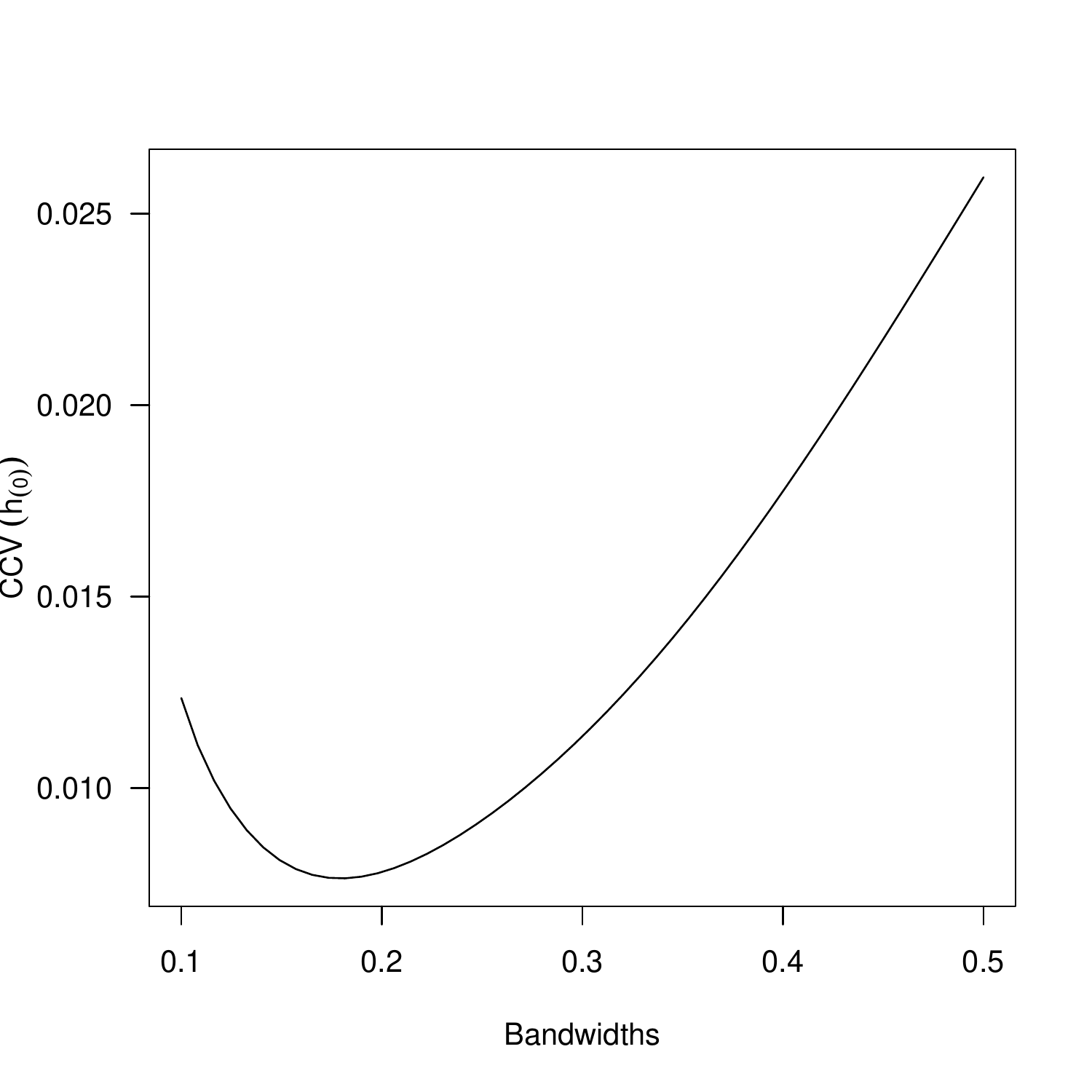}
\includegraphics{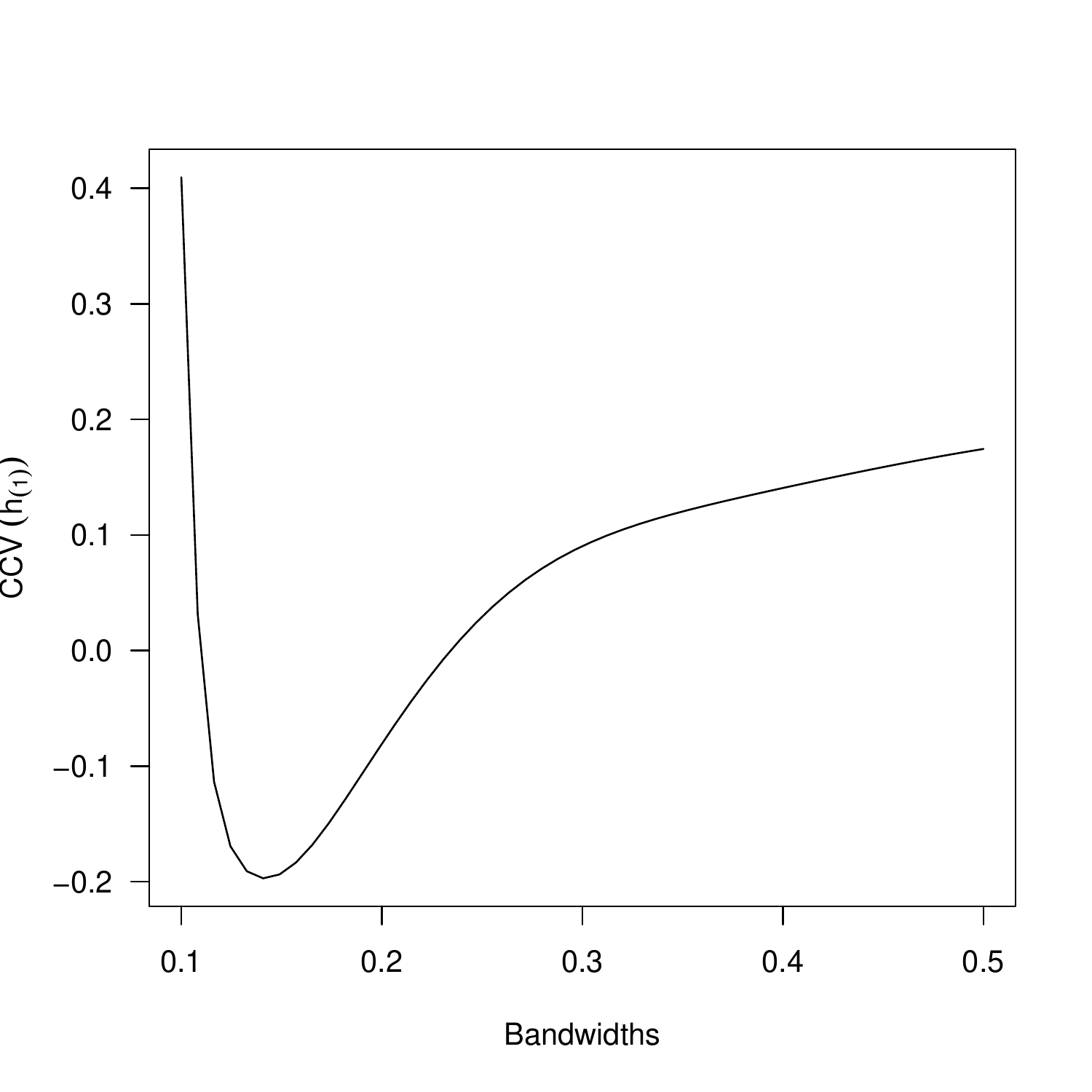}
\includegraphics{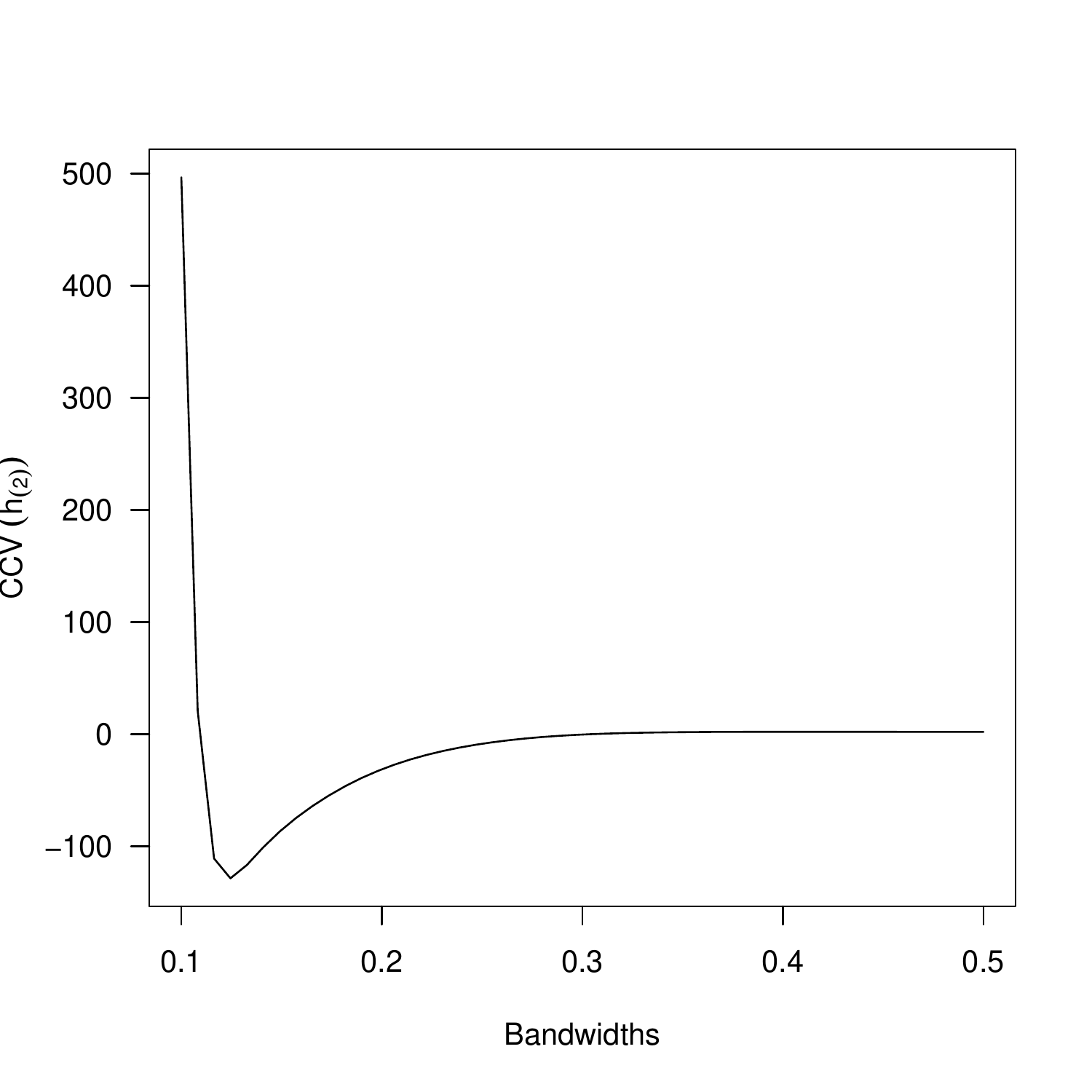}
\includegraphics{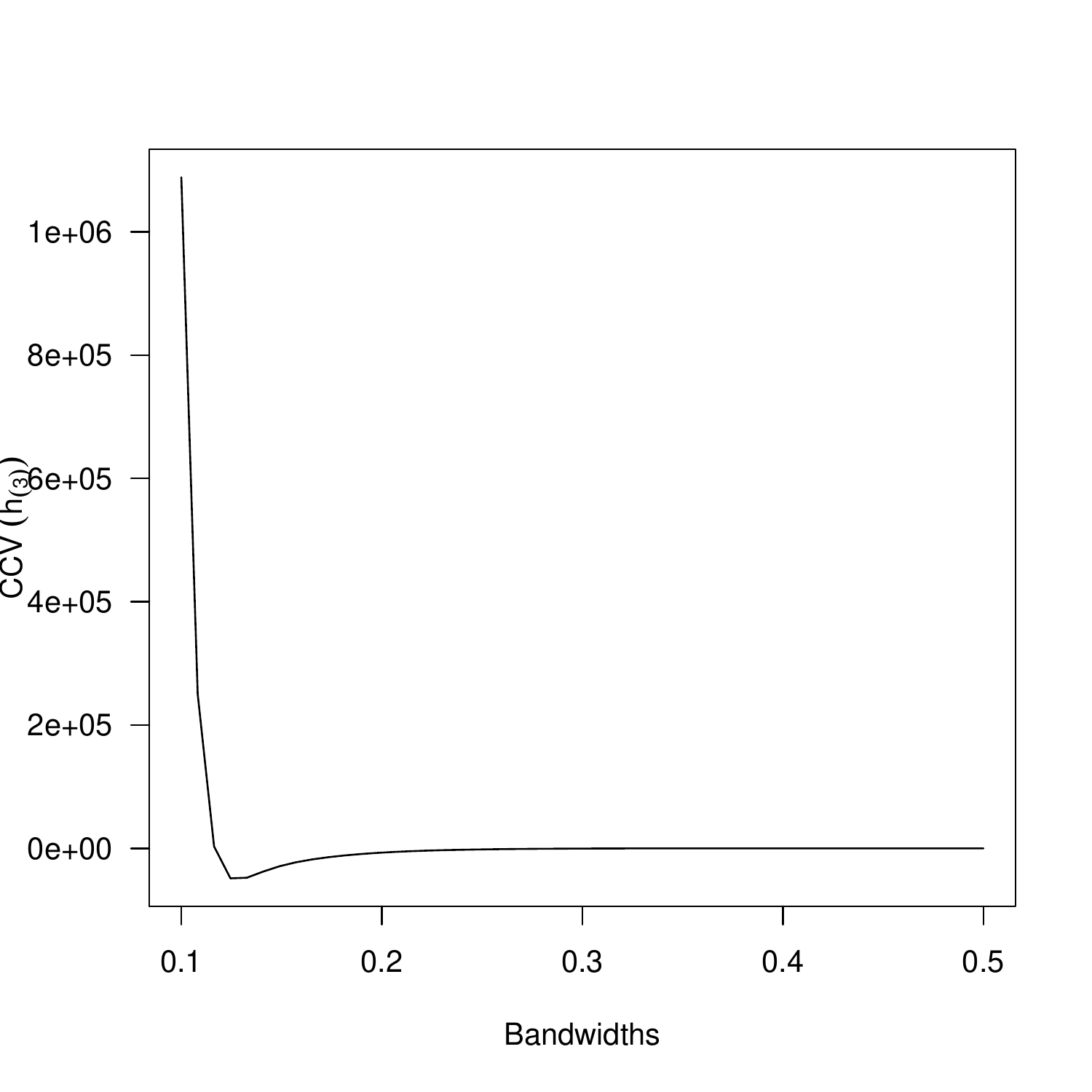}
\end{center}
\caption{$\CCV$ function obtained by \code{h.ccv()}. (top left) \code{deriv.order = 0}. (top right) \code{deriv.order = 1}. (bottom left) \code{deriv.order = 2}. (bottom right) \code{deriv.order = 3}.}\label{Sec3:fig4}
\end{figure}

The following example computes the bandwidth $h$ by this method for a first three derivatives estimators of \eqref{Sec2:eq2}. This time we set Over-smoothing in \code{upper = 0.5}.
\begin{Schunk}
\begin{Sinput}
R> h.ccv(bimodal, deriv.order = 0, upper = 0.5)
\end{Sinput}
\begin{Soutput}
Call:		Complete Cross-Validation

Derivative order = 0
Data: bimodal (200 obs.);	Kernel: gaussian
Min CCV = 0.00764383;	Bandwidth 'h' = 0.1790795
\end{Soutput}
\begin{Sinput}
R> h.ccv(bimodal, deriv.order = 1, upper = 0.5)
\end{Sinput}
\begin{Soutput}
Call:		Complete Cross-Validation

Derivative order = 1
Data: bimodal (200 obs.);	Kernel: gaussian
Min CCV = -0.197061;	Bandwidth 'h' = 0.1427856
\end{Soutput}
\begin{Sinput}
R> h.ccv(bimodal, deriv.order = 2, upper = 0.5)
\end{Sinput}
\begin{Soutput}
Call:		Complete Cross-Validation

Derivative order = 2
Data: bimodal (200 obs.);	Kernel: gaussian
Min CCV = -127.6184;	Bandwidth 'h' = 0.1259258
\end{Soutput}
\begin{Sinput}
R> h.ccv(bimodal, deriv.order = 3, upper = 0.5)
\end{Sinput}
\begin{Soutput}
Call:		Complete Cross-Validation

Derivative order = 3
Data: bimodal (200 obs.);	Kernel: gaussian
Min CCV = -35547.32;	Bandwidth 'h' = 0.1426445
\end{Soutput}
\end{Schunk}
The plot of $\CCV$  function obtained with the code (Figure \ref{Sec3:fig4}):
\begin{Schunk}
\begin{Sinput}
R> for (i in 0:3) 
+  plot(h.ccv(bimodal, deriv.order = i), seq.bws=seq(0.1,0.5,length=50))
\end{Sinput}
\end{Schunk}
\setkeys{Gin}{width=0.45\textwidth}

\subsection{Modified cross-validation}

\cite{Stute1992} proposed a so-called modified cross-validation ($\MCV$) in kernel density estimator. This method can be extended to the estimation of derivative of a probability density, the essential idea based on approximated the problematic term by the aid of the Hajek projection. The minimization criterion is defined by:
\begin{equation}\label{Sec3:eq11}
  \MCV(h,r) =  \frac{\RK}{nh^{2r+1}} + \frac{(-1)^r}{n(n-1) h^{2r+1}} \Sum2 \varphi^{(r)} \Z
\end{equation}
where$$ \varphi^{(r)} (c) = \left(\ConvKr - K^{(2r)} - \frac{\M2(K)}{2} K^{(2r+2)} \right)(c)$$
The function \code{h.mcv()} computes the modified cross-validation for bandwidth selection. We enumerate the arguments and results of this function in Table \ref{Sec3:Tab6}.

\begin{table}[!ht]
	\centering
\begin{tabular}{ll}
  \toprule
  Arguments & Description \\
  \midrule
  \code{x} &  The data sample.\\
  \code{deriv.order} & Derivative order (scalar). \\
  \code{lower,upper} &  Range over which to minimize. The default is almost always satisfactory,\\
                     &  \code{hos} (Over-smoothing) is calculated internally from an kernel.\\
  \code{tol} &  The convergence tolerance for optimize.\\
  \code{kernel} & The kernel function (see Table \ref{Sec1:Tab1}), by default \code{"gaussian"}. \\
   \midrule\midrule
  Results & Description \\
  \midrule\midrule
  \code{h} & Value of bandwidth.\\
  \code{min.mcv} & The minimal $\MCV$ value (Equation \ref{Sec3:eq11}).\\
  \bottomrule
\end{tabular}
\caption{Summary of arguments and results of \code{h.mcv()}.}\label{Sec3:Tab6}
\end{table}

\begin{figure}[!ht]
\begin{center}
\includegraphics{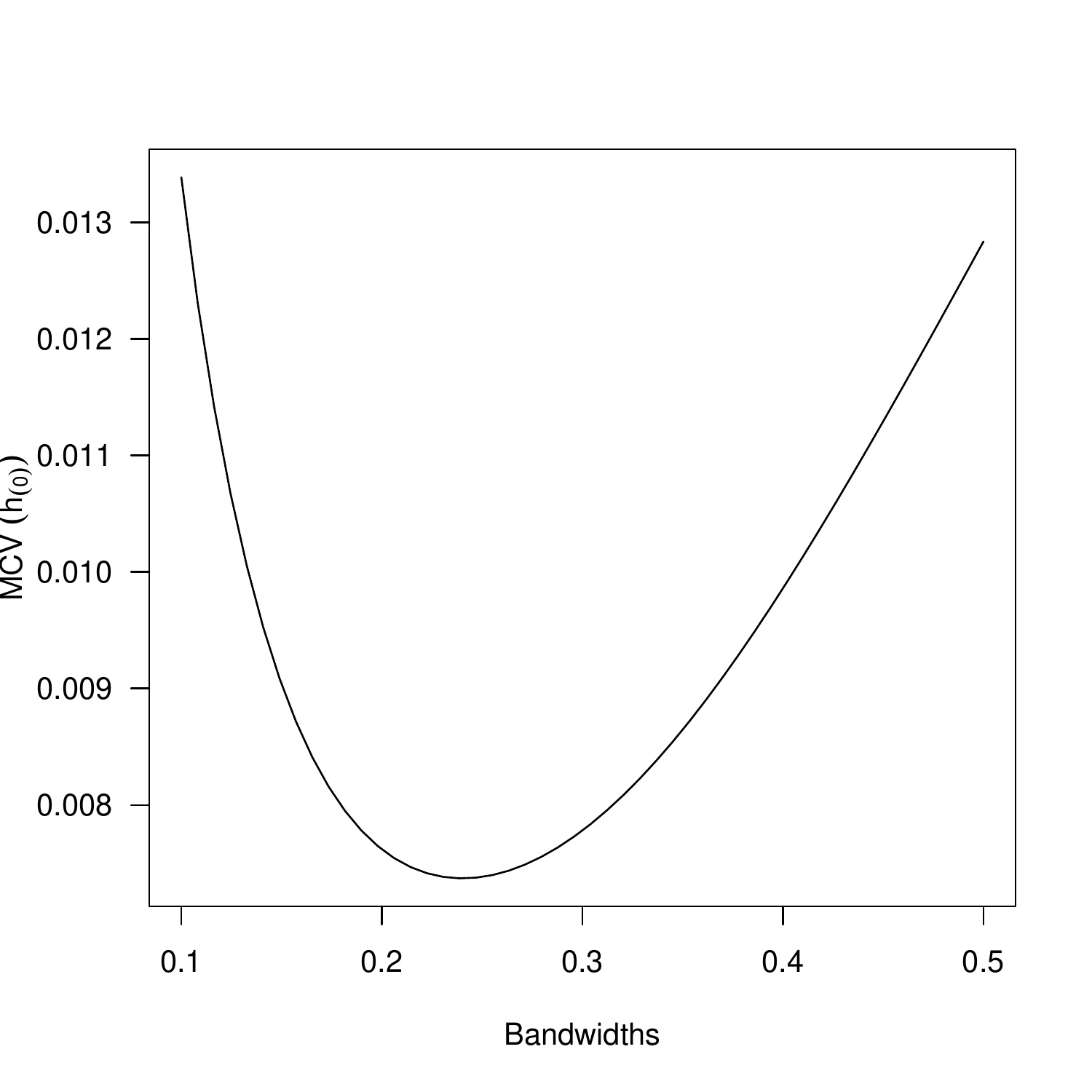}
\includegraphics{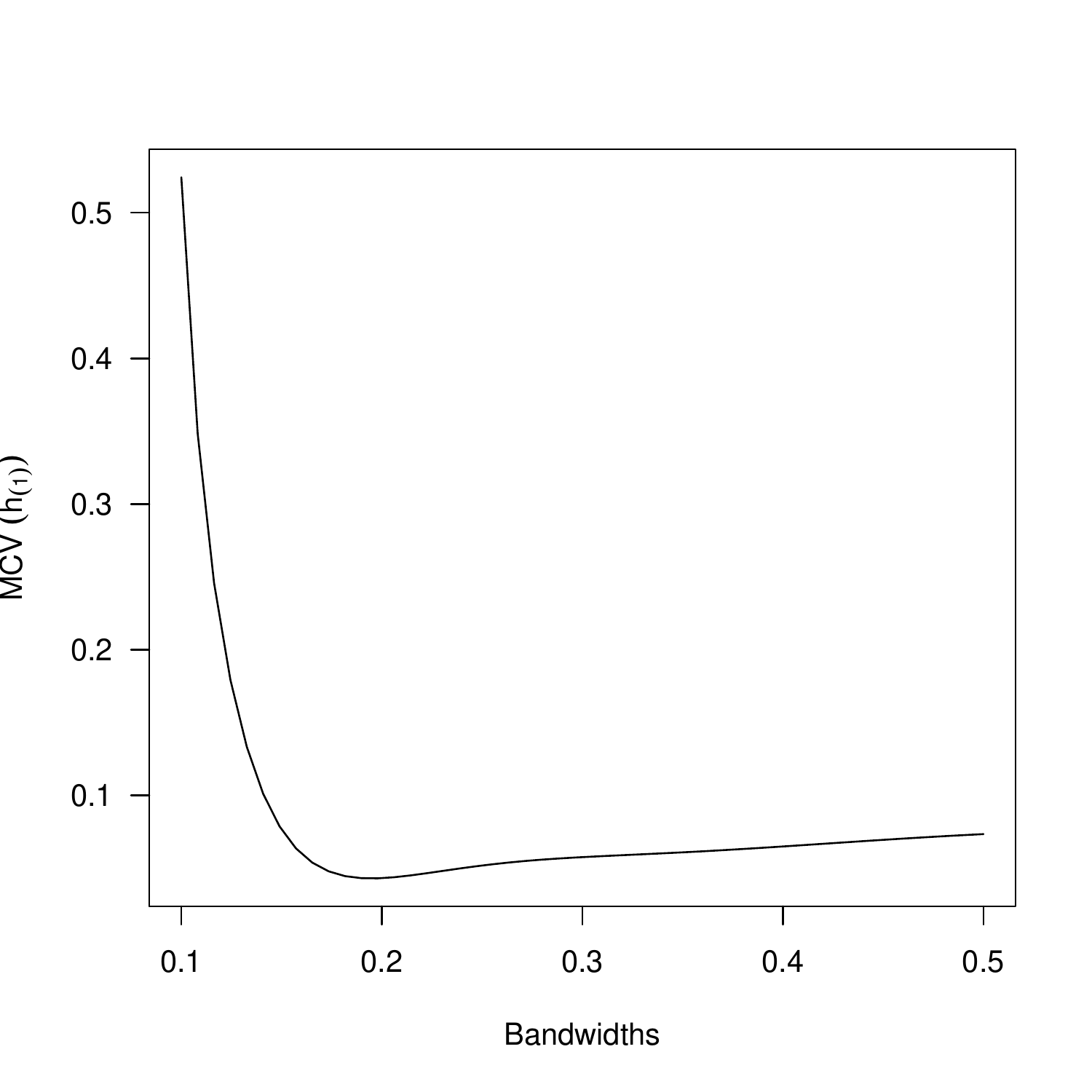}
\includegraphics{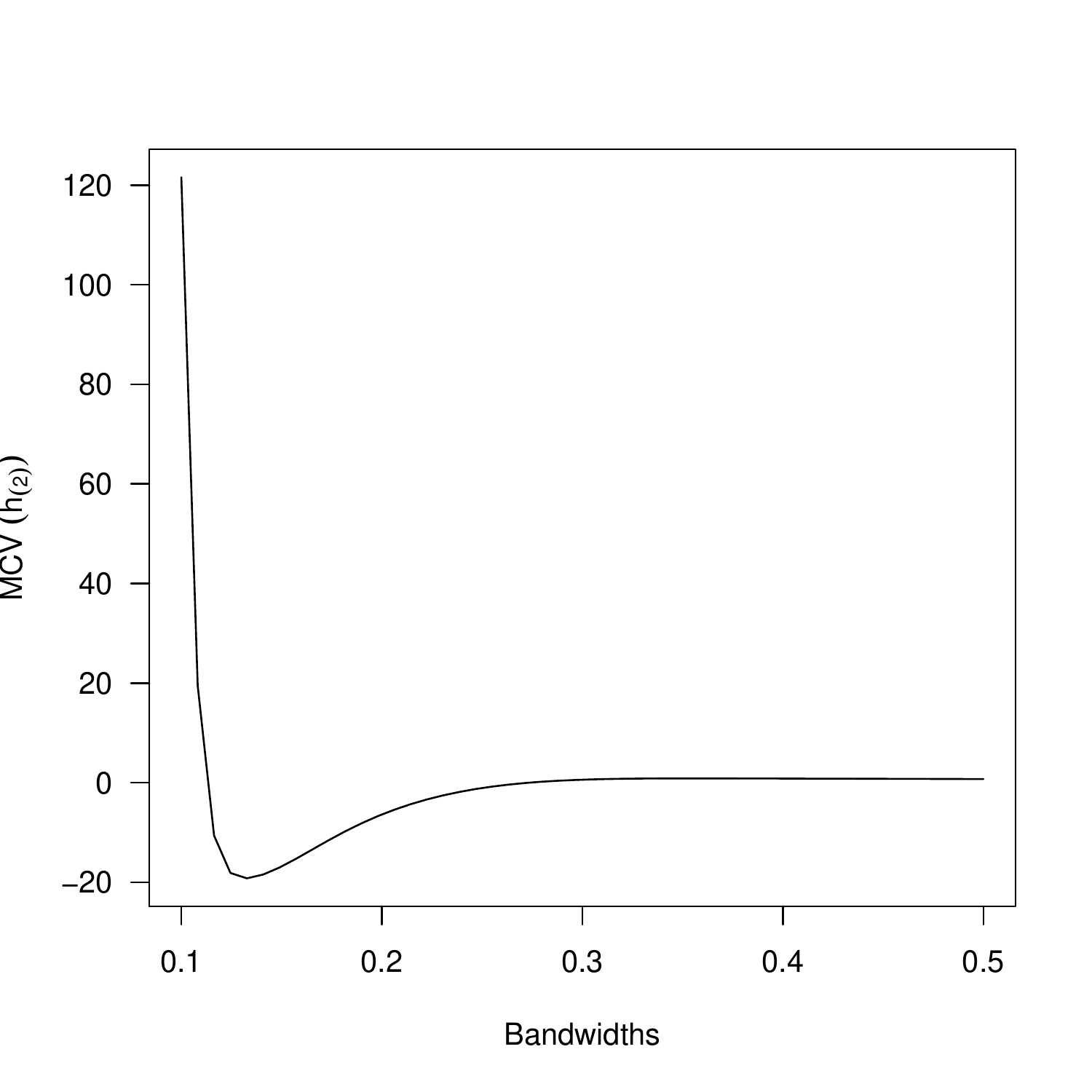}
\includegraphics{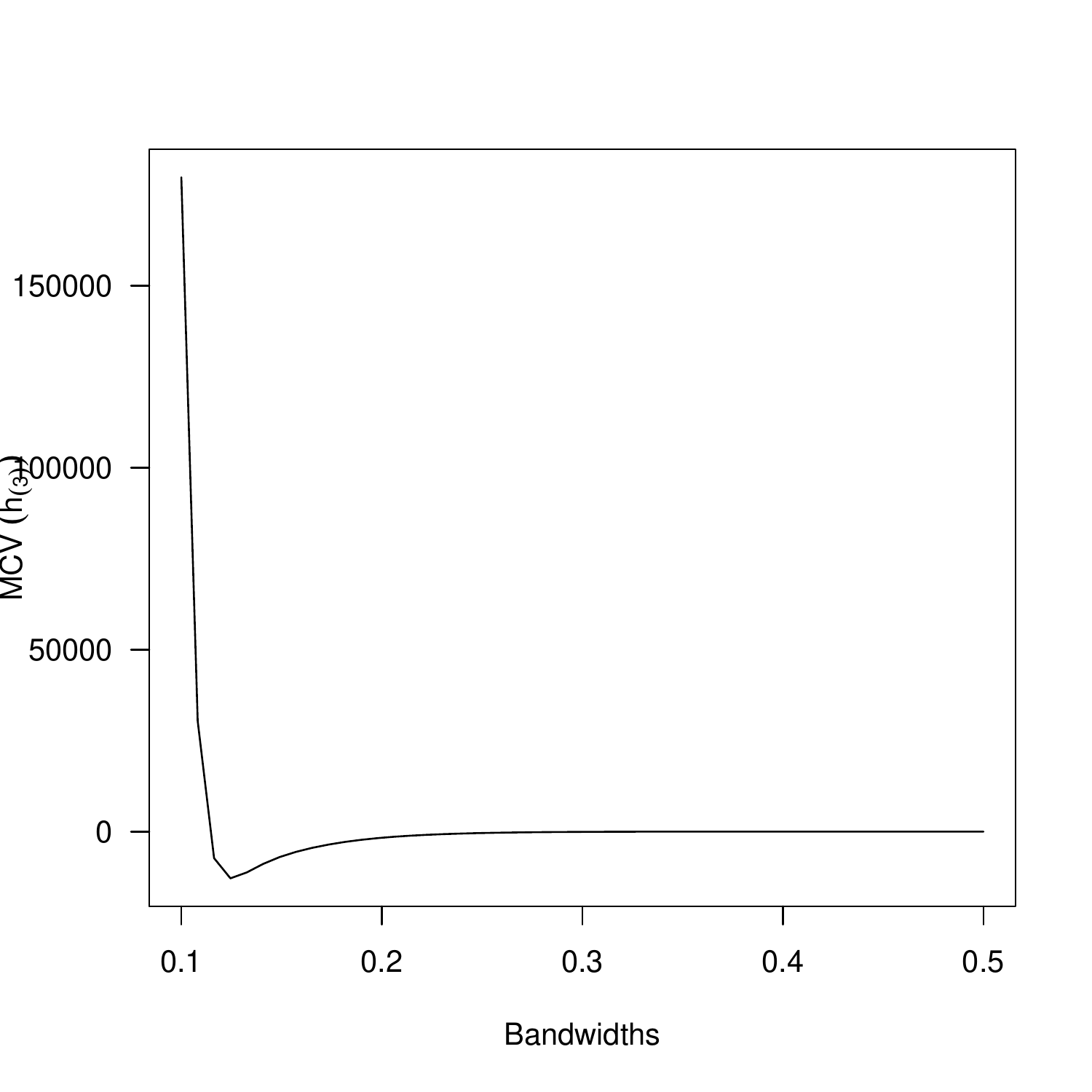}
\end{center}
\caption{$\MCV$ function obtained by \code{h.mcv()}. (top left) \code{deriv.order = 0}. (top right) \code{deriv.order = 1}. (bottom left) \code{deriv.order = 2}. (bottom right) \code{deriv.order = 3}.}\label{Sec3:fig5}
\end{figure}

The following example computes the bandwidth $h$ by this method for a first three derivatives estimators of \eqref{Sec2:eq2}. We set Over-smoothing in \code{upper = 0.5}.
\begin{Schunk}
\begin{Sinput}
R> h.mcv(bimodal, deriv.order = 0, upper = 0.5)
\end{Sinput}
\begin{Soutput}
Call:		Modified Cross-Validation

Derivative order = 0
Data: bimodal (200 obs.);	Kernel: gaussian
Min MCV = 0.007370888;	Bandwidth 'h' = 0.2402603
\end{Soutput}
\begin{Sinput}
R> h.mcv(bimodal, deriv.order = 1, upper = 0.5)
\end{Sinput}
\begin{Soutput}
Call:		Modified Cross-Validation

Derivative order = 1
Data: bimodal (200 obs.);	Kernel: gaussian
Min MCV = 0.04308772;	Bandwidth 'h' = 0.196102
\end{Soutput}
\begin{Sinput}
R> h.mcv(bimodal, deriv.order = 2, upper = 0.5)
\end{Sinput}
\begin{Soutput}
Call:		Modified Cross-Validation

Derivative order = 2
Data: bimodal (200 obs.);	Kernel: gaussian
Min MCV = -19.17733;	Bandwidth 'h' = 0.1304198
\end{Soutput}
\begin{Sinput}
R> h.mcv(bimodal, deriv.order = 3, upper = 0.5)
\end{Sinput}
\begin{Soutput}
Call:		Modified Cross-Validation

Derivative order = 3
Data: bimodal (200 obs.);	Kernel: gaussian
Min MCV = -8391.447;	Bandwidth 'h' = 0.1426445
\end{Soutput}
\end{Schunk}
The plot of $\MCV$  function obtained with the code (Figure \ref{Sec3:fig5}):
\begin{Schunk}
\begin{Sinput}
> for (i in 0:3)
+  plot(h.mcv(bimodal, deriv.order = i), seq.bws=seq(0.1,0.5,length=50))
\end{Sinput}
\end{Schunk}
\setkeys{Gin}{width=0.45\textwidth}

\subsection{Trimmed cross-validation}

\cite{FeluchandKoronacki1992} proposed a so-called trimmed cross-validation ($\TCV$) in kernel density estimator, a simple modification of the unbiased (least-squares) cross-validation criterion \eqref{Sec3:eq7}. We consider the following "trimmed" version of "unbiased", to be minimized with respect to $h$:
\begin{equation}\label{Sec3:eq12}
    \TCV(h,r) =  \frac{\RK}{nh^{2r+1}} + \frac{(-1)^r}{n(n-1) h^{2r+1}} \Sum2 \varphi^{(r)} \Z
\end{equation}
where$$ \varphi^{(r)} (c) = \left[\ConvKr - 2 K^{(2r)} 1\left(|c| > \frac{c_{n}}{h^{2r+1}}\right) \right](c)$$
$1(.)$ denotes the indicator function and $c_{n}$ is a sequence of positive constants, as $\lim_{n \to \infty} c_{n}/h \rightarrow 0$, here we take $c_{n} = 1/n$, for assure the convergence.

The function \code{h.tcv()} computes the trimmed cross-validation for bandwidth selection. We enumerate the arguments and results of this function in Table \ref{Sec3:Tab7}.

\begin{table}[!ht]
		\centering
\begin{tabular}{ll}
 \toprule
  Arguments & Description \\
  \midrule
  \code{x} &  The data sample.\\
  \code{deriv.order} & Derivative order (scalar). \\
  \code{lower,upper} &  Range over which to minimize. The default is almost always satisfactory,\\
                     &  \code{hos} (Over-smoothing) is calculated internally from an kernel.\\
  \code{tol} &  The convergence tolerance for optimize.\\
  \code{kernel} & The kernel function (see Table \ref{Sec1:Tab1}), by default \code{"gaussian"}. \\
   \midrule\midrule
  Results & Description \\
  \midrule\midrule
  \code{h} & Value of bandwidth.\\
  \code{min.tcv} & The minimal $\TCV$ value (Equation \ref{Sec3:eq12}).\\
  \bottomrule
\end{tabular}
\caption{Summary of arguments and results of \code{h.tcv()}.}\label{Sec3:Tab7}
\end{table}

\begin{figure}[!ht]
\begin{center}
\includegraphics{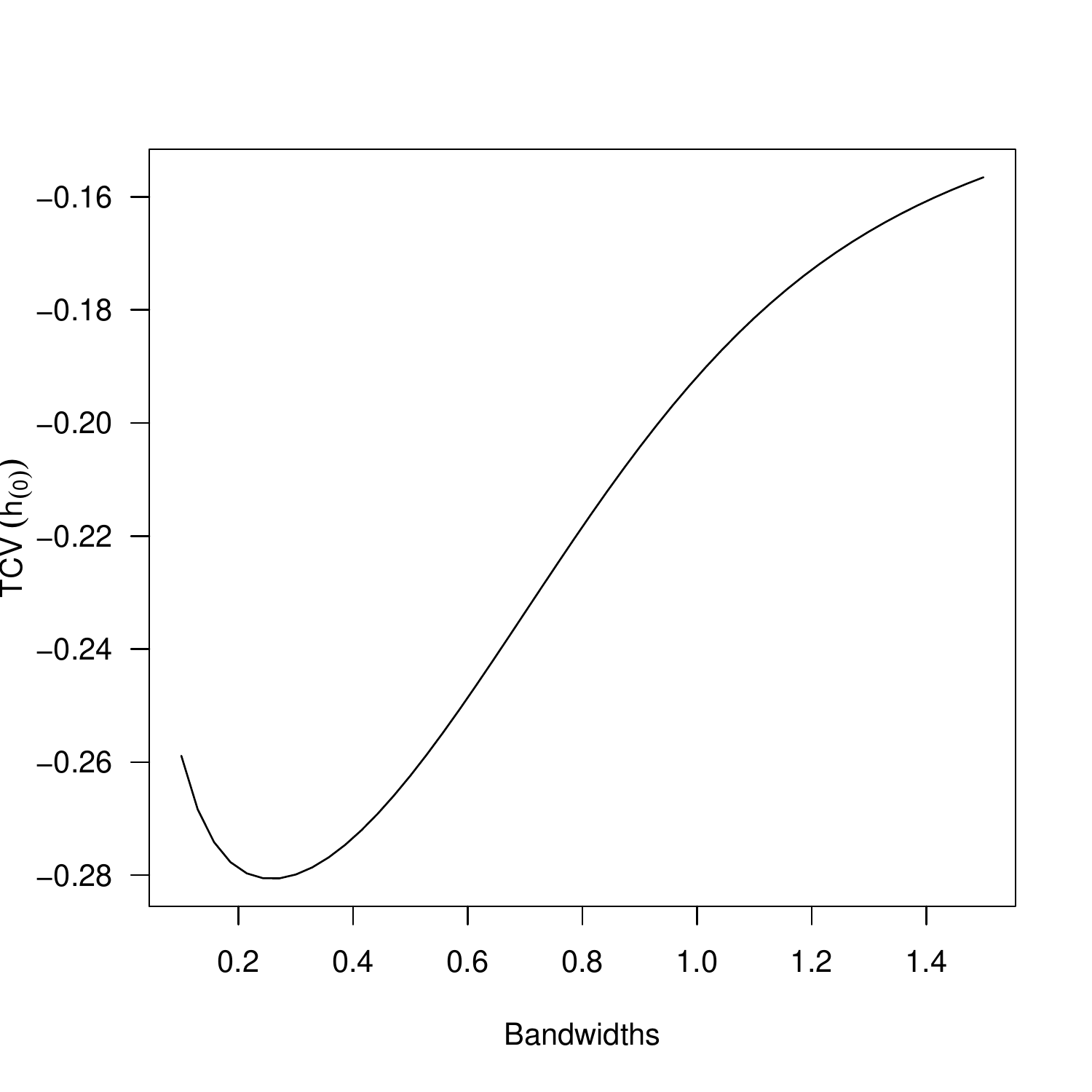}
\includegraphics{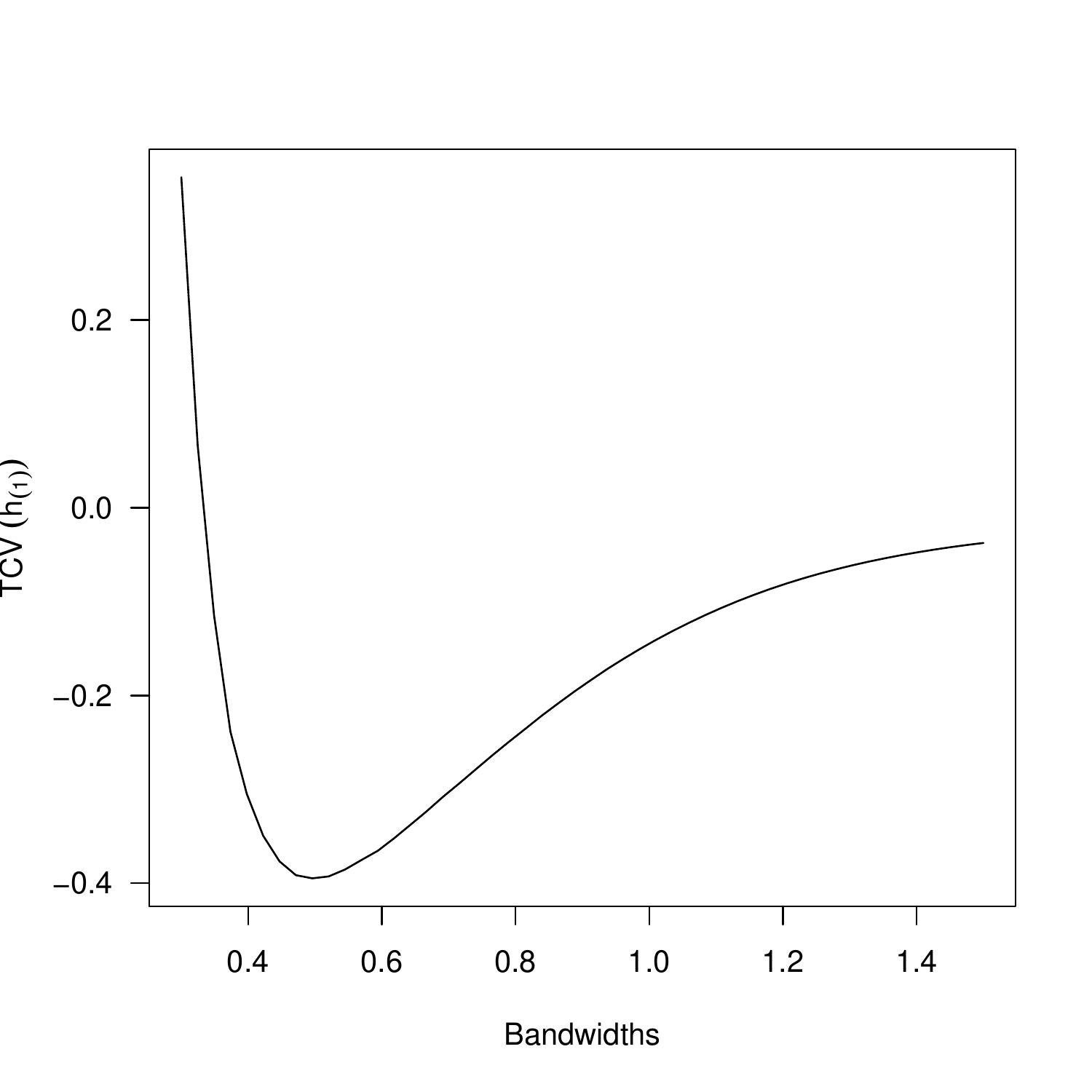}
\includegraphics{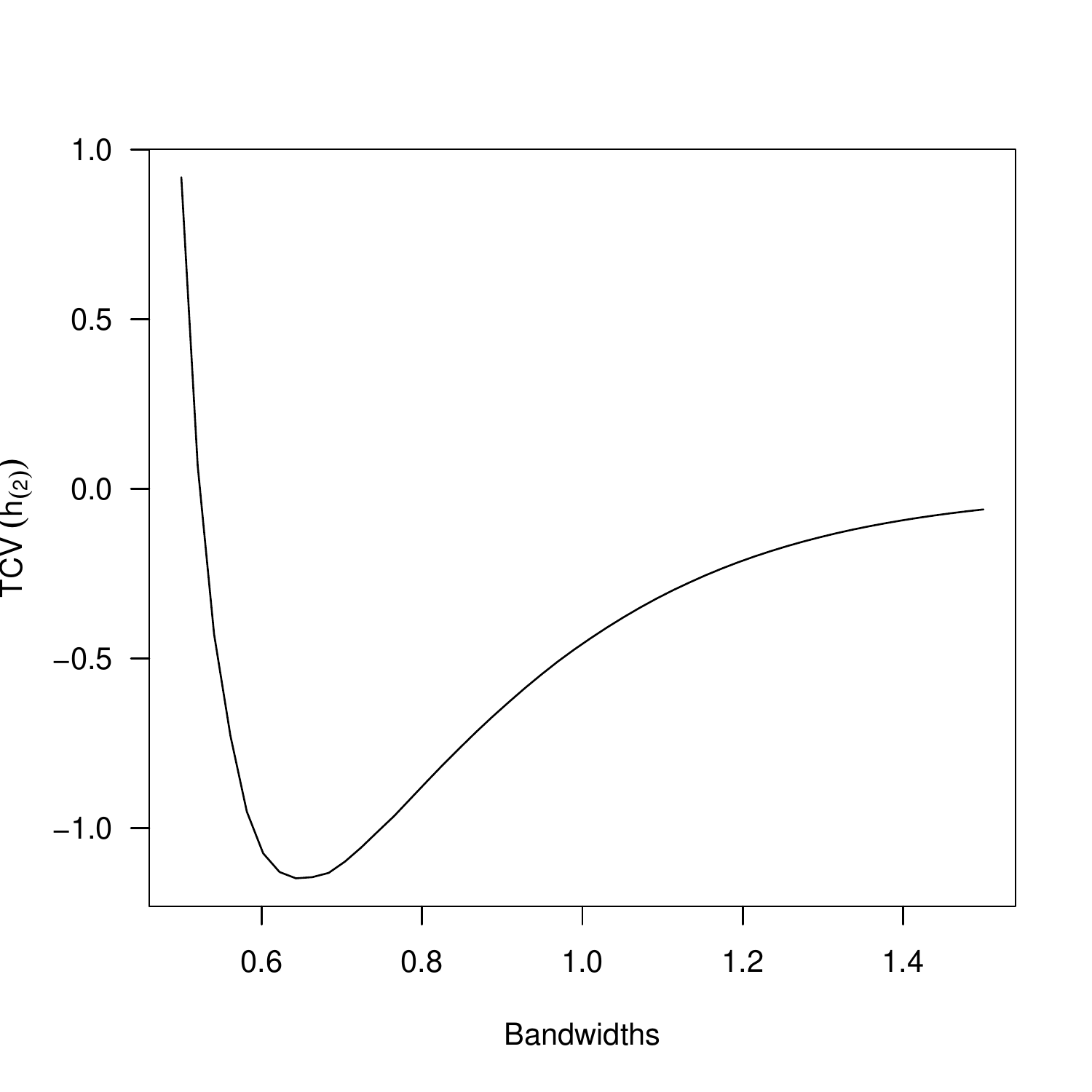}
\includegraphics{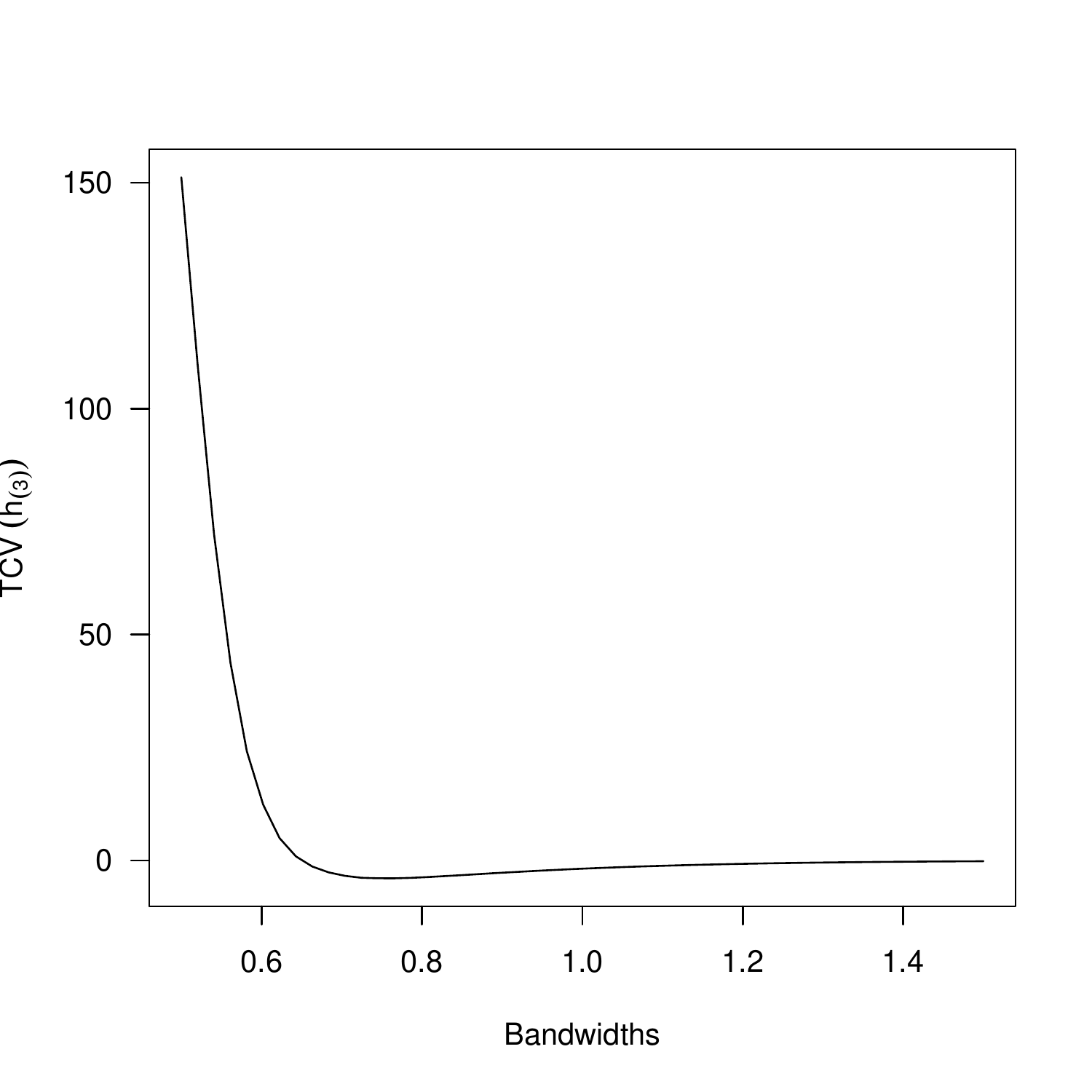}
\end{center}
\caption{$\TCV$ function obtained by \code{h.tcv()}. (top left) \code{deriv.order = 0}. (top right) \code{deriv.order = 1}. (bottom left) \code{deriv.order = 2}. (bottom right) \code{deriv.order = 3}.}\label{Sec3:fig6}
\end{figure}

The following example computes the bandwidth $h$ by this method for a first three derivatives estimators of \eqref{Sec2:eq2}.
\begin{Schunk}
\begin{Sinput}
R> h.tcv(bimodal, deriv.order = 0)
\end{Sinput}
\begin{Soutput}
Call:		Trimmed Cross-Validation

Derivative order = 0
Data: bimodal (200 obs.);	Kernel: gaussian
Min TCV = -0.2806542;	Bandwidth 'h' = 0.2570739
\end{Soutput}
\begin{Sinput}
R> h.tcv(bimodal, deriv.order = 1)
\end{Sinput}
\begin{Soutput}
Call:		Trimmed Cross-Validation

Derivative order = 1
Data: bimodal (200 obs.);	Kernel: gaussian
Min TCV = -0.3956968;	Bandwidth 'h' = 0.5095463
\end{Soutput}
\begin{Sinput}
R> h.tcv(bimodal, deriv.order = 2)
\end{Sinput}
\begin{Soutput}
Call:		Trimmed Cross-Validation

Derivative order = 2
Data: bimodal (200 obs.);	Kernel: gaussian
Min TCV = -1.153544;	Bandwidth 'h' = 0.65599
\end{Soutput}
\begin{Sinput}
R> h.tcv(bimodal, deriv.order = 3)
\end{Sinput}
\begin{Soutput}
Call:		Trimmed Cross-Validation

Derivative order = 3
Data: bimodal (200 obs.);	Kernel: gaussian
Min TCV = -3.961823;	Bandwidth 'h' = 0.7453813
\end{Soutput}
\end{Schunk}
The plot of $\TCV$  function obtained with the code (Figure \ref{Sec3:fig6}):
\begin{Schunk}
\begin{Sinput}
R> for (i in 0:3) plot(h.tcv(bimodal, deriv.order = i))
\end{Sinput}
\end{Schunk}
\setkeys{Gin}{width=0.45\textwidth}
\vspace{3cm}
\section{Summary}

We have implemented in \texttt{R} the estimators of the defined functions and the bandwidth selection procedures of the above sections. The package \pkg{kedd} contains seven functions, in Table \ref{Sec4:Tab1} we can find a summary of the contents of the package. The current feature set of the package can be split in four main categories: compute the convolutions and derivatives of a kernel function, compute the kernel estimators for a density of probability and its derivatives, computing the bandwidth selectors with different methods, displaying the kernel estimators and selection functions of the bandwidth. Moreover, the package follows the general \texttt{R} philosophy of working with model objects. This means that instead of merely returning, say, a kernel estimator of $\rth$ derivative of a density, many functions will return an object containing, it's functions are S3 classes (\code{S3method}). The object can then be manipulated at will using various extraction, summary or plotting functions. Whenever possible, we develop a graphical user interface of the various functions of a coherent whole, to facilitate the use of this package.
\begin{table}[!ht]
\centering
\begin{center}
\begin{tabular}{ll}
  \toprule
  Function & Description \\
  \midrule
  \code{dkde} &  Derivatives of kernel density estimator, as defined in Equation \ref{Sec2:eq1}.\\
  \code{h.amise} & $\AMISE$ for optimal bandwidth selectors (Equation \ref{Sec3:eq3}). \\
  \code{h.mlcv} &  Maximum-likelihood cross-validation bandwidth selection (Equation \ref{Sec3:eq5}).\\
  \code{h.ucv} &  Unbiased cross-validation bandwidth selection (Equation \ref{Sec3:eq7}).\\
  \code{h.bcv} &  Biased cross-validation bandwidth selection (Equations \ref{Sec3:eq8} and \ref{Sec3:eq9}) .\\
  \code{h.ccv} &  Complete cross-validation bandwidth selection (Equation \ref{Sec3:eq10}).\\
  \code{h.mcv} &  Modified cross-validation bandwidth selection (Equation \ref{Sec3:eq11}).\\
  \code{h.tcv} &  Trimmed cross-validation bandwidth selection (Equation \ref{Sec3:eq12}).\\
  \bottomrule
\end{tabular}
\end{center}
\caption{Summary of contents of the package.}\label{Sec4:Tab1}
\end{table}

\bibliographystyle{unsrt}

\begin{thebibliography}{2}

\bibitem[Alekseev, 1972]{Alekseev}
Alekseev, V. G. (1972).
\newblock  Estimation of a probability density function and its derivatives.
\newblock  \emph{Mathematical notes of the Academy of Sciences of the USSR}. \textbf{12}(5), 808--811.

\bibitem[Alexandre, 2009]{Alexandre}
Alexandre, B. T. (2009).
\newblock \emph{Introduction to Nonparametric Estimation}.
\newblock Springer-Verlag, New York.

\bibitem[Bhattacharya, 1967]{Bhattacharya}
Bhattacharya, P. K. (1967).
\newblock  Estimation of a probability density function and Its derivatives.
\newblock  \emph{Sankhya: The Indian Journal of Statistics, Series A}, \textbf{29}, 373--382.

\bibitem[Bowman, 1984]{Bowman1984}
Bowman, A. W. (1984).
\newblock  An alternative method of cross-validation for the smoothing of kernel density estimates.
\newblock  \emph{Biometrika}, \textbf{71}, 353--360.

\bibitem[Bowman and Azzalini, 1997]{Bowman}
Bowman, A. W. and Azzalini, A. (1997).
\newblock  \emph{Applied Smoothing Techniques for Data Analysis: the Kernel Approach with S-Plus Illustrations}.
\newblock  Oxford University Press, Oxford.

\bibitem[Bowman and Azzalini, 2013]{smarticle}
Bowman, A. W. and Azzalini, A. (2013).
\newblock \texttt{R} package \pkg{sm}: nonparametric smoothing methods (version 2.2-5).
\newblock \url{http://www.stats.gla.ac.uk/~adrian/sm, http://azzalini.stat.unipd.it/Book_sm}

\bibitem[Chiu, 1991a]{Chiu1991a}
Chiu, S.T. (1991a).
\newblock Some stabilized bandwidth selectors for nonparametric regression.
\newblock \emph{Ann. Stat.} \textbf{19}, 1528--1546.

\bibitem[Chiu, 1991b]{Chiu1991b}
Chiu, S.T. (1991b).
\newblock Bandwidth selection for kernel density estimation.
\newblock \emph{Ann. Stat.} \textbf{19}, 1883--1905.

\bibitem[Chiu, 1992]{Chiu1992}
Chiu, S.T. (1992).
\newblock An automatic bandwidth selector for kernel density estimation.
\newblock \emph{Biometrika}, \textbf{79}, 771--782.

\bibitem[Duong, 2007]{ks}
Duong, T. (2007).
\newblock \pkg{ks}: {K}ernel density estimation and kernel discriminant analysis for multivariate data in \texttt{R}.
\newblock {\em Journal of Statistical Software}. \textbf{21}(7).

\bibitem[Duong and Hazelton, 2005]{DuongandHazelton2005}
Duong, T. and Hazelton, M.L. (2005).
\newblock Cross-validation bandwidth matrices for multivariate kernel density estimation.
\newblock \emph{Scandinavian Journal of Statistics}, \textbf{32}, 485--506.

\bibitem[Duong and Hazelton, 2003]{DuongandHazelton2003}
Duong, T. and Hazelton, M.L. (2003).
\newblock Plug-in bandwidth selectors for bivariate kernel density estimation.
\newblock \emph{Journal of Nonparametric Statistics}, \textbf{15}, 17--30.

\bibitem[Duong and Matt, 2013]{feature}
Duong, T. and Matt, W. (2013).
\newblock \pkg{feature}: Feature significance for multivariate kernel density estimation.
\newblock \texttt{R} package version 1.2.9.
\newblock \url{http://CRAN.R-project.org/package=feature}

\bibitem[Duin (1976)]{Duin1976}
Duin, R. P. W. (1976).
\newblock  On the choice of smoothing parameters of Parzen estimators of probability density functions.
\newblock  \emph{IEEE Transactions on Computers}, \textbf{C-25}, 1175--1179.

\bibitem[Feluch and Koronacki, 1992]{FeluchandKoronacki1992}
Feluch, W. and Koronacki, J. (1992).
\newblock  A note on modified cross-validation in density estimation.
\newblock  \emph{Computational Statistics and Data Analysis}, \textbf{13}, 143--151.

\bibitem[Guidoum, 2015]{kedd}
Guidoum, A. C. (2015).
\newblock \pkg{kedd}: Kernel estimator and bandwidth selection for density and its derivatives.
\newblock \texttt{R} package version 1.0.3.
\newblock \url{http://CRAN.R-project.org/package=kedd}

\bibitem[Habbema, Hermans and Van den Broek (1974)]{Habbema1974}
Habbema, J. D. F., Hermans, J., and Van den Broek, K. (1974).
\newblock  A stepwise discrimination analysis program using density estimation.
\newblock \emph{Compstat 1974: Proceedings in Computational Statistics}. Physica Verlag, Vienna.

\bibitem[Heidenreich et all, 2013]{Heidenreichetall2013}
Heidenreich, N. B., Schindler, A. and Sperlich, S. (2013).
\newblock  Bandwidth selection for kernel density estimation: a review of fully automatic selectors.
\newblock  \emph{Advances in Statistical Analysis}.

\bibitem[Jeffrey, 1996]{Jeffrey}
Jeffrey, S. S. (1996).
\newblock  \emph{Smoothing Methods in Statistics}.
\newblock  Springer-Verlag, New York.

\bibitem[Jones and Kappenman, 1991]{JonesandKappenman1991}
Jones, M. C. and Kappenman, R. F. (1991).
\newblock  On a class of kernel density estimate bandwidth selectors.
\newblock  \emph{Scandinavian Journal of Statistics}, \textbf{19}, 337--349.

\bibitem[Jones et all, 1996]{Jonesetall1996}
Jones, M. C., Marron, J. S. and Sheather,S. J. (1996).
\newblock  A brief survey of bandwidth selection for density estimation.
\newblock  \emph{Journal of the American Statistical Association}, \textbf{91}, 401--407.

\bibitem[Olver et all, 2010]{Olver}
Olver, F. W., Lozier, D. W., Boisvert, R. F. and Clark, C. W. (2010).
\newblock  \emph{NIST Handbook of Mathematical Functions}.
\newblock  Cambridge University Press, New York, USA.

\bibitem[Peter and Marron, 1987]{PeterandMarron1987}
Peter, H. and Marron, J.S. (1987).
\newblock  Estimation of integrated squared density derivatives.
\newblock  \emph{Statistics and Probability Letters}, \textbf{6}, 109--115.

\bibitem[Peter and Marron, 1991]{PeterandMarron1991}
Peter, H. and Marron, J.S. (1991).
\newblock  Local minima in cross-validation functions.
\newblock  \emph{Journal of the Royal Statistical Society, Series B}, \textbf{53}, 245--252.

\bibitem[\texttt{R} Development Core Team (2015)]{R}
R Development Core Team (2015).
\newblock {\em \texttt{R}: A Language and Environment for Statistical Computing}.
\newblock Vienna, Austria.
\newblock \url{http://www.R-project.org/}

\bibitem[Rudemo, 1982]{Rudemo1982}
Rudemo, M. (1982).
\newblock   Empirical choice of histograms and kernel density estimators.
\newblock   \emph{Scandinavian Journal of Statistics}, \textbf{9}, 65--78.

\bibitem[Schuster, 1969]{Schuster}
Schuster, E. F. (1969).
\newblock  Estimation of a probability density function and its derivatives.
\newblock  \emph{The Annals of Mathematical Statistics}, \textbf{40}(4), 1187--1195.

\bibitem[Scott, 1992]{Scott1992}
Scott, D. W. (1992).
\newblock \emph{Multivariate Density Estimation. Theory, Practice and Visualization}.
\newblock New York: Wiley.

\bibitem[Scott and George, 1987]{ScottandGeorge1987}
Scott, D.W. and George, R. T. (1987).
\newblock  Biased and unbiased cross-validation in density estimation.
\newblock  \emph{Journal of the American Statistical Association}, \textbf{82}, 1131--1146.

\bibitem[Sheather, 2004]{Sheather2004}
Sheather, S. J. (2004).
\newblock  Density estimation.
\newblock  \emph{Statistical Science}, \textbf{19}, 588--597.

\bibitem[Sheather and Jones, 1991]{SheatherandJones1991}
Sheather, S. J. and Jones, M. C. (1991).
\newblock  A reliable data-based bandwidth selection method for kernel density estimation.
\newblock  \emph{Journal of the Royal Statistical Society, Series B}, \textbf{53}, 683--690.

\bibitem[Silverman, 1986]{Silverman}
Silverman, B. W. (1986).
\newblock \emph{Density Estimation for Statistics and Data Analysis}.
\newblock Chapman \& Hall/CRC. London.

\bibitem[Singh, 1977]{Singh1977}
Singh, R. S. (1990).
\newblock  Applications of estimators of a density and its derivatives\textbf{39}(3), 357--363.

\bibitem[Stoker, 1993]{Stoker1993}
Stoker, T. M. (1993).
\newblock  Smoothing bias in density derivative estimation.
\newblock  \emph{Journal of the American Statistical Association}, \textbf{88}, 855--863.

\bibitem[Stute, 1992]{Stute1992}
Stute, W. (1992).
\newblock  Modified cross validation in density estimation.
\newblock  \emph{Journal of Statistical Planning and Inference}, \textbf{30}, 293--305.

\bibitem[Tristen and Jeffrey, 2008]{np}
Tristen, H. and Jeffrey, S. R. (2008).
\newblock Nonparametric Econometrics: The \pkg{np} Package.
\newblock {\em Journal of Statistical Software}. \textbf{27 (5)}.

\bibitem[Wand and Jones, 1995]{WandandJones}
Wand, M. P. and Jones, M. C. (1995).
\newblock  \emph{Kernel Smoothing}.
\newblock  Chapman and Hall, London.

\bibitem[Wand and Ripley, 2013]{KernSmooth}
Wand, M.P. and Ripley, B. D. (2013).
\newblock \pkg{KernSmooth}: Functions for kernel smoothing for Wand and Jones (1995).
\newblock \texttt{R} package version 2.23-10.
\newblock \url{http://CRAN.R-project.org/package=KernSmooth}

\bibitem[Wolfgang, 1991]{Wolfgang}
Wolfgang, H. (1991).
\newblock  \emph{Smoothing Techniques, With Implementation in S}.
\newblock  Springer-Verlag, New York.

\bibitem[Wolfgang et all, 1990]{Wolfgangetall1990}
Wolfgang, H., Marron, J. S. and Wand, M. P. (1990).
\newblock  Bandwidth choice for density derivatives.
\newblock  \emph{Journal of the Royal Statistical Society, Series B}, 223--232.

\bibitem[Wolfgang et all, 2004]{Wolfgangetall}
Wolfgang, H., Marlene, M., Stefan, S. and Axel, W. (2004).
\newblock  \emph{Nonparametric and Semiparametric Models}.
\newblock  Springer-Verlag, Berlin Heidelberg.

\bibitem[Venables and Ripley, 2002]{VenablesandRipley}
Venables, W. N. and Ripley, B. D. (2002).
\newblock  \emph{Modern Applied Statistics with S}.
\newblock  New York: Springer.

\end{thebibliography}

\end{document}